\documentclass[aps,prd,english,nofrontpage,floatfix,nofootinbib,superscriptaddress,aps,prd,showkeys]{revtex4}
\usepackage[utf8]{inputenc}
\usepackage{float}
\usepackage{array}
\usepackage{lipsum}
\usepackage{graphicx}
\usepackage{amsmath,amsthm,amsfonts,amssymb}
\usepackage{tikz}
\usepackage{multirow}

\newcommand{\be}{\begin{equation}}
\newcommand{\ee}{\end{equation}}
\newcommand{\Be}{\begin{eqnarray}}
\newcommand{\Ee}{\end{eqnarray}}
\newcommand{\f}{\frac}
\newcommand{\pa}{\partial}

\newcommand{\mincir}{\raise
-3.truept\hbox{\rlap{\hbox{$\sim$}}\raise4.truept\hbox{$<$}\ }}
\newcommand{\magcir}{\raise
-3.truept\hbox{\rlap{\hbox{$\sim$}}\raise4.truept\hbox{$>$}\ }}

\newcolumntype{Y}{>{\centering\arraybackslash}X}
\providecommand{\U}[1]

 \usetikzlibrary{quotes,angles}
 \usetikzlibrary{arrows} 
\usetikzlibrary{decorations.markings}
\usepackage{graphicx}
\usepackage[english]{babel}
\usepackage{color}
\usepackage{subfig}
\usepackage{caption}
\usepackage{tensor}
\usepackage{esint}
\usepackage[dvips]{epsfig}
\usepackage[dvips]{graphicx}
\usepackage{float}
\usepackage{units}
\usepackage{textcomp}
\usepackage{mathrsfs}
\usepackage{amsmath}
\usepackage[makeroom]{cancel}
\usepackage{amssymb}
\usepackage{amsbsy}
\usepackage{amsfonts}
\usepackage{amssymb,mathrsfs,xcolor}
\usepackage{esint}
\usepackage{array}
\usepackage{graphicx}

\usepackage{hyperref}
\hypersetup{
    colorlinks,
    citecolor=blue,
    filecolor=green,
    linkcolor=purple,
    urlcolor=red,
}

\usepackage{slashed}

\newcommand{\ie}{\begin{equation}}
\newcommand{\fe}{\end{equation}}
\newcommand{\se}{\begin{eqnarray}}
\newcommand{\ff}{\end{eqnarray}}

\usepackage{hyperref}
\hypersetup{colorlinks,breaklinks,
			citecolor=[rgb]{0,0.0,1.0},
            urlcolor=[rgb]{0.0,0.0,1.0},
            linkcolor=[rgb]{0,0.5,0.9}}

\begin{document}

\title{Effects of non-commutative geometry on black hole properties}

\author{A. A. Ara\'{u}jo Filho}
\email{dilto@fisica.ufc.br}

\author{J. R. Nascimento}
\email{jroberto@fisica.ufpb.br}
\affiliation{Departamento de Física, Universidade Federal da Paraíba, Caixa Postal 5008, 58051-970, João Pessoa, Paraíba,  Brazil.}

\author{A. Yu. Petrov}
\email{petrov@fisica.ufpb.br}
\affiliation{Departamento de Física, Universidade Federal da Paraíba, Caixa Postal 5008, 58051-970, João Pessoa, Paraíba,  Brazil.}

\author{P. J. Porfírio}
\email{pporfirio@fisica.ufpb.br}
\affiliation{Departamento de Física, Universidade Federal da Paraíba, Caixa Postal 5008, 58051-970, João Pessoa, Paraíba,  Brazil.} 

\author{Ali \"Ovg\"un
}
\email{ali.ovgun@emu.edu.tr}
\affiliation{Physics Department, Eastern Mediterranean
University, Famagusta, 99628 North Cyprus, via Mersin 10, Turkiye}

\date{\today}

\begin{abstract}

In this study, we investigate the signatures of a non-commutative black hole solution. Initially, we calculate the thermodynamic properties of the system, including entropy, heat capacity, and \textit{Hawking} radiation. For the latter quantity, we employ two distinct methods: surface gravity and the topological approach. Additionally, we examine the emission rate and remnant mass within this context. Remarkably, the lifetime of the black hole, after reaching its final state due to the evaporation process, is expressed analytically up to a \textit{grey-body} factor. We estimate the lifetime for specific initial and final mass configurations. Also, we analyze the tensorial quasinormal modes using the 6th-order WKB method. Finally, we study the deflection angle, i.e., gravitational lensing, in both the weak and strong deflection limits.

%the gravitational lensing of a non-commutative background through gravitational gauge field potentials, emphasizing a static spherically symmetric black hole. Our analysis employs the strong deflection limit for calculating the deflection angle and addressing the lens equation, yielding measurable outcomes such as relativistic image positions and magnifications. Also, we explore the shadows as well as the emission rate in this context. Finally, we propose a practical application of our findings in understanding galactic phenomena associated with Sgr A*.

\end{abstract}


\maketitle

%\tableofcontents
\section{Introduction}
\label{sec:intro}

In the domain of general relativity, the formalism used to depict spacetime geometry lacks a precise boundary on distance measurements, often regarding the Planck length as a foundational constraint. To address this challenge, scholars commonly explore the concept of non-commutative spacetimes. The motivation for exploring non-commutative geometry stems from its connections to string/M-theory \cite{szabo2006symmetry,szabo2003quantum,3}, with notable applications observed in the domain of supersymmetric Yang-Mills theories, particularly within the superfield framework \cite{ferrari2003finiteness,ferrari2004superfield,ferrari2004towards}. Additionally, the Seiberg-Witten map serves as a common tool for introducing non-commutativity into gravitational theories by gauging an appropriate group \cite{chamseddine2001deforming}. This framework has led to significant advancements in understanding black holes, encompassing their dynamics of evaporation \cite{myung2007thermodynamics,t29} and thermodynamic characteristics \cite{banerjee2008noncommutative,lopez2006towards,sharif2011thermodynamics,nozari2006reissner,nozari2007thermodynamics}.

Nicolini et al. \cite{nicolini2006noncommutative} introduced a significant discovery regarding the non-commutative effect. They proposed a novel model that alters the matter source term while preserving the Einstein tensor part of the field equation. In this model, the conventional point-like mass density on the Einstein equation is substituted by either a Gaussian \cite{ghosh2018noncommutative} smeared distribution or a Lorentzian distribution \cite{nicolini2009noncommutative}. These distributions are defined as ${\rho _\Theta } = M{(4\pi \Theta )^{ - \frac{3}{2}}}e^{- \frac{{{r^2}}}{{4\Theta }}}$ and ${\rho _\Theta } = M\sqrt \Theta {\pi ^{ - \frac{3}{2}}}{({r^2} + \pi \Theta )^{ - 2}}$, respectively. In these contexts, numerous studies have explored various aspects of black hole physics, including the \textit{Hawking} temperature and tunneling effects \cite{nozari2008hawking,banerjee2008noncommutative,sharif2011thermodynamics}, shadow behavior \cite{ovgun2020shadow,sharif2016shadow,wei2015shadow}, topological features in Gauss-Bonnet gravity \cite{lekbich2024optical}, gravitational lensing \cite{ding2011strong,ding2011probing}, and accretion of matter \cite{saleem2023observable}. Moreover, a novel approach for incorporating non-commutativity into gravitational scenarios has emerged in the literature, fundamentally treating it as a perturbation \cite{newcommutativity}

Gravitational waves and their characteristics are crucial for understanding a diverse array of physical phenomena, ranging from early universe events to astrophysical occurrences such as stellar oscillations and binary systems \cite{unno1979nonradial, kjeldsen1994amplitudes, dziembowski1992effects, pretorius2005evolution, hurley2002evolution, yakut2005evolution, heuvel2011compact}. These waves exhibit diverse intensities and modes, with their spectral attributes intricately linked to their sources \cite{riles2017recent}. Notably, the emission of gravitational waves from black holes holds significant importance. Following the collapse of matter, BHs emit radiation characterized by unique frequencies termed \textit{quasinormal} modes \cite{rincon2020greybody, santos2016quasinormal, oliveira2019quasinormal, berti2009quasinormal, horowitz2000quasinormal, nollert1999quasinormal, ferrari1984new, jusufi2024charged,heidari2024impact,kokkotas1999quasi, araujo2024dark,london2014modeling, maggiore2008physical, flachi2013quasinormal, ovgun2018quasinormal, blazquez2018scalar, roy2020revisiting, konoplya2011quasinormal,Hamil:2024ppj}. In the literature, it is widely employed the weak field approximation method to study these modes within black hole contexts, encompassing general relativity and other gravitational theories, as well as scenarios involving Lorentz violation and related fields \cite{kim2018quasi, lee2020quasi, jawad2020quasinormal,araujo2024exploring,maluf2013matter, maluf2014einstein, JCAP1, JCAP2, JCAP3, jcap4, jcap5,hassanabadi2023gravitational,Heidari:2023bww,mmm1,mmm2,gogoi2023quasinormal,liu2022quasinormal,yang2023probing,Gogoi:2023fow,lambiase2023investigating,Fernando:2016ftj,Fernando:2012yw,Daghigh:2008jz,Daghigh:2020jyk,Daghigh:2005ph,Daghigh:2020mog,Daghigh:2020fmw,Daghigh:2022uws,Gogoi:2023kjt,Yang:2022ifo}.

Recent advancements, particularly the detection of gravitational waves by the LIGO-Virgo collaboration \cite{016,017,018}, have expanded the scope of cosmological research. Gravitational waves are now used to explore the universe, including studying gravitational lensing within the weak field approximation \cite{019,020,vagnozzi2022horizon}. Historically, gravitational lensing studies focused on light traveling great distances from the gravitational source, such as in Schwarzschild spacetime \cite{021}, and were later extended to general spherically symmetric and static spacetimes \cite{022}. However, in regions with strong gravitational fields, like those near black holes, the angular deviation of light is greatly amplified.

Observations from the Event Horizon Telescope of a supermassive black hole at the center of the M87 galaxy have generated significant scientific interest \cite{023,024,025,026,027,028,029,pantig2023testing,ccimdiker2021black,lambiase2023investigating,pantig2022shadow}. Early work by Virbhadra and Ellis introduced a concise lens equation for supermassive black holes within an asymptotically flat background \cite{030,031}, showing multiple symmetrically distributed images around the optic axis due to strong gravitational effects. Further advancements made by Fritelli et al. \cite{032}, Bozza et al. \cite{033}, and Tsukamoto \cite{035} have enhanced the analytical frameworks for studying strong field gravitational lensing. These studies have examined light deflection in various contexts, including Reissner-Nordström spacetime \cite{036,036.1,036.2}, rotating solutions \cite{37.1,37.2,37.3,37.4,37.5,37.6}, exotic constructs like wormholes \cite{38.1,38.2,38.3,38.4,38.5}, and modified gravity theories \cite{40}.

In this study, we explore the characteristics of a non-commutative black hole solution. We begin by calculating the system's thermodynamic properties, including entropy, heat capacity, and \textit{Hawking} radiation. For the latter, we use two distinct methods: surface gravity and the topological approach. Additionally, we investigate the emission rate and remnant mass in this context. Notably, the black hole's lifetime, post-evaporation, is analytically expressed up to a \textit{grey-body} factor, with estimations made for specific initial and final mass configurations. We also analyze the tensorial quasinormal modes using the 6th-order WKB method. Lastly, we examine the deflection angle, or gravitational lensing, in both the weak and strong deflection limits.

The structure of this work is as follows: In Sec. \ref{general}, we introduce the general framework of the non-commutative theory under consideration, including the mass distribution, black hole solution, and horizons. In Sec. \ref{tee}, we discuss the thermodynamics, evaporation, and emission rate. Sec. \ref{tee} also covers the calculation of the \textit{quasinormal} modes using tensorial perturbations. In Sec. \ref{weak}, we present gravitational lensing calculations in the weak deflection angle limit via the Gauss-Bonnet theorem. In Sec. \ref{strong}, we revisit gravitational lensing using the strong deflection limit technique. Finally, in Sec. \ref{conclu}, we offer our concluding remarks.

%%%%%%%%%%%%%%%%%%%%%%%%%%%%%%%%%%%%%%%%%%%%%%%%%%%%%%%%%%%%%%%%%%%%%%%%%%%%%%%%%%%%%%%%%%%%%%%%%%%%%%%%%%%%%%%%%%%%%%%%%%%%%%%%%%%%%%%%%%%%%%%%%%%%%%%%%%%%%%%%%%%%%%%%%%%%%%%%%%%%%%%%%%%%%%%%%%%%%%%%%%%%%%%%%%%%%%%%%%%%%%%%%%%%%%%%%%%%%%%%%%%%%%%%%%%%%%%%%%%%%%%%%%%%%%%%%%%%%%%%%%%%%%%%%%%%%%%%%%%%%%%%%%%%%%%%%%%%%%%%%%%%%%%%%%%%%%%%%%%%%%%%%%%%%%%%%%%%%%%%%%%%%%%%%%%%%%%%%%%%%%%%%%%%%%%%%%%%%%%%%%%%%%%%%%%%%%%%%%%%%%%%%%%%%%

\section{The general setup\label{general}}

\begin{figure}
    \centering
     \includegraphics[scale=0.42]{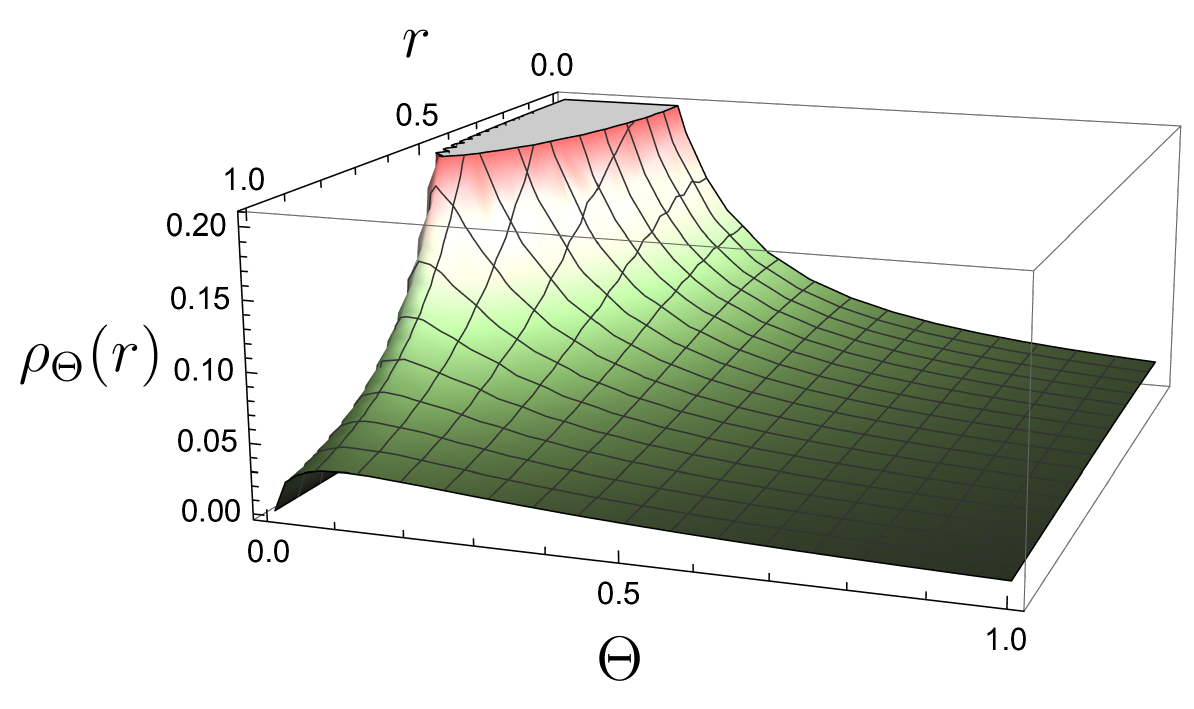}
    \caption{The representation of the distribution $\rho_{\Theta}(r)$.}
    \label{distribution}
\end{figure}

\begin{figure}
    \centering
     \includegraphics[scale=0.4]{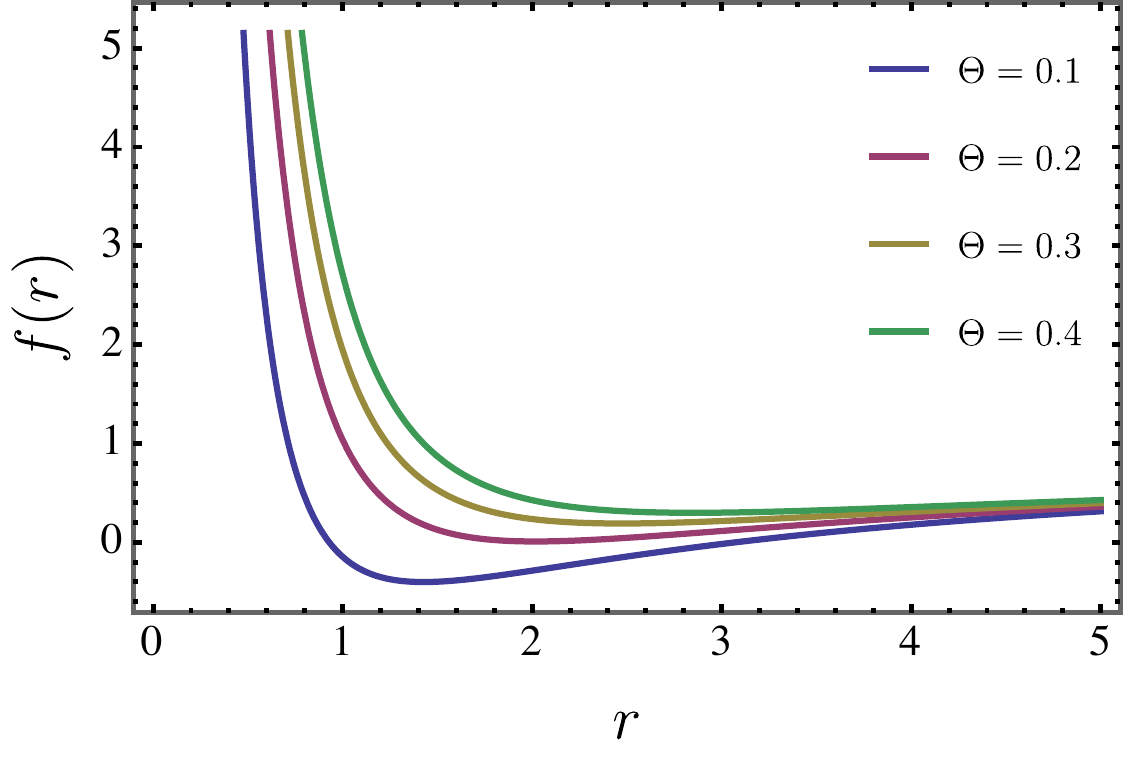}
    \caption{The representation of $f(r)$ for different values of $\Theta$.}
    \label{horizons}
\end{figure}

Examining the implications of spacetime involves, for instance, taking into account non-commutativity principles with general relativity \cite{k10,k101,k102,k103,k6,k7,k8,k9,campos2022quasinormal}. Various formulations of non-commutative field theory, based on the Moyal product, have been proposed \cite{k11}. This section initiates by examining key features of the black hole solution in question, beginning with the provided distribution \cite{campos2022quasinormal,nicolini2006noncommutative,nozari2008hawking}
\ie
\rho_{\Theta}(r) = \frac{M \sqrt{\Theta}}{\pi^{3/2} (r^{2} + \pi \Theta)^{2}},
\fe
where $M$ is the total mass and $\Theta$ is the non-commutative parameter with dimension of $[\mathrm{L}^{2}]$, which is defined as
\ie
[x^{\mu},x^{\nu}] = i \Theta^{\mu \nu}.
\fe
In Fig. \ref{distribution}, we represent the distribution $\rho_{\Theta}(r)$, considering different values of the non-commutative parameter $\Theta$. Here, we can also define $\mathcal{M}_{\Theta}$ as
\ie
\mathcal{M}_{\Theta} = \int^{r}_{0} 4\pi r^{2}\rho_{\Theta}(r) \mathrm{d}r = M - \frac{4M \sqrt{\Theta}}{\sqrt{\pi}r}.
\fe
In possession of it, in the non-commutative scenario, the  Schwarzschild-like black hole is
\ie
\label{lalalaa}
\mathrm{d}s^{2} = -f_{\Theta} (r) \mathrm{d}t^{2} + f_{\Theta} (r) \mathrm{d}r^{2} + r^{2}\mathrm{d}\theta^{2} + r^{2}\sin^{2}\theta \mathrm{d}\varphi^{2},
\fe
in which 
\ie
f_{\Theta}(r) = 1- \frac{2M}{r} + \frac{8M\sqrt{\Theta}}{\sqrt{\pi}r^{2}}.
\fe
Such a metric gives rise to two physical solutions
\ie
r_{+} = M + \frac{\sqrt{\pi  M^2-8 \sqrt{\pi } \sqrt{\Theta } M}}{\sqrt{\pi }},
\fe
and
\ie
r_{-} = M-\frac{\sqrt{\pi  M^2-8 \sqrt{\pi } \sqrt{\Theta } M}}{\sqrt{\pi }},
\fe
where $r_{+}$ and $r_{-}$ are the event the Cauchy horizon respectively. To enhance reader comprehension, we present plots and tables illustrating both quantities. Fig. \ref{horizons} depicts $r_{+}$ and $r_{-}$ as functions of $M$ for various values of $\Theta$. Furthermore, Tab. \ref{evenntthorizon} reveals that as $M$ increases, so does $r_{+}$, while an increase in $\Theta$ corresponds to a decrease in the radius of the event horizon. Conversely, Tab. \ref{couchyhorizonzz} demonstrates that as mass increases, $r_{-}$ decreases; additionally, an increase in the non-commutative parameter results in a growth of the radius of the Cauchy horizon.

\begin{figure}
    \centering
     \includegraphics[scale=0.4]{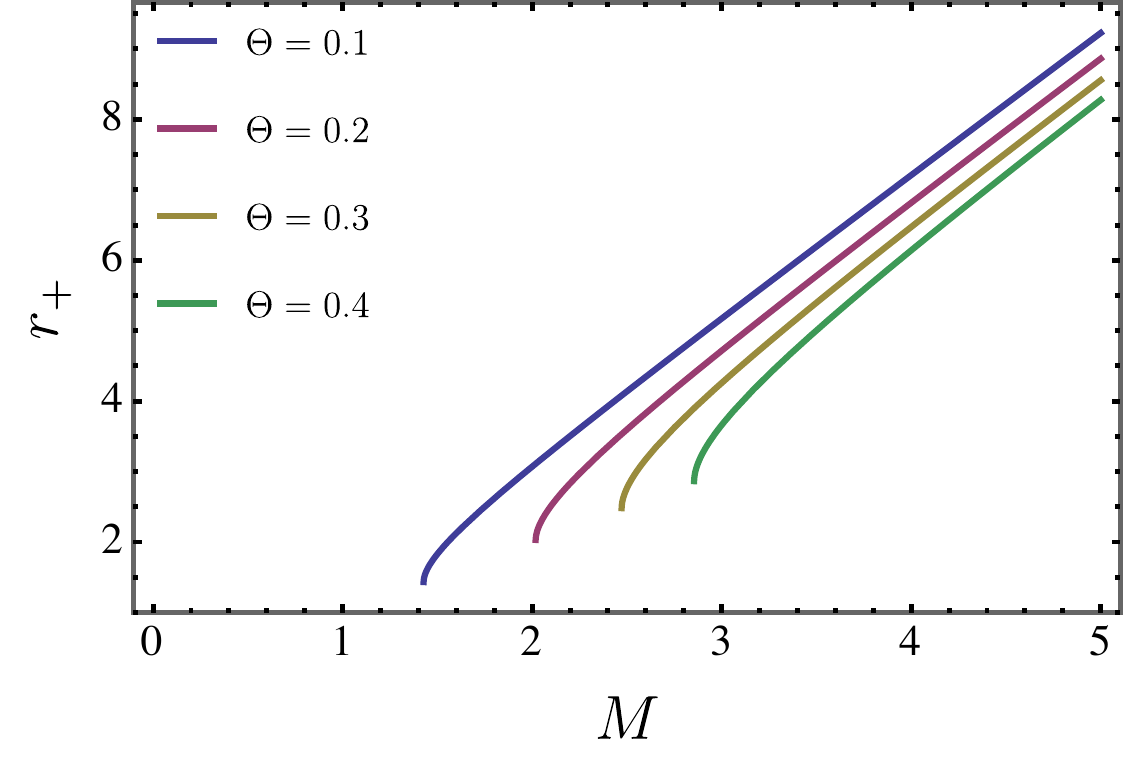}
     \includegraphics[scale=0.4105]{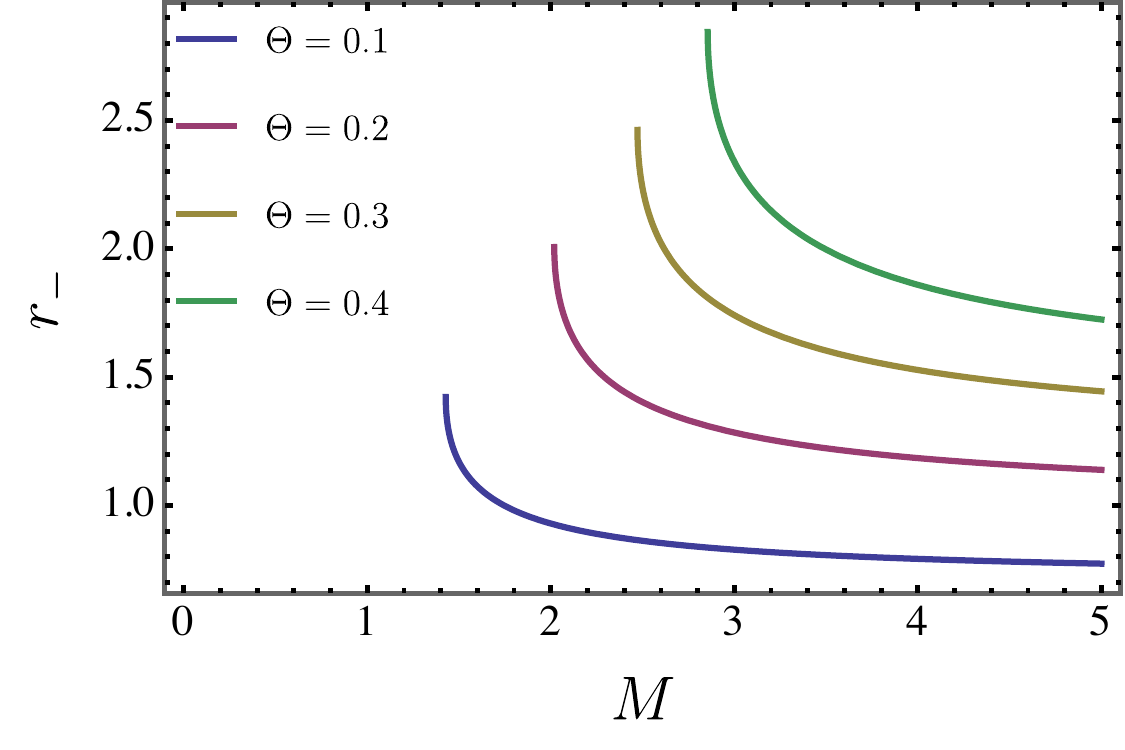}
    \caption{The event (on the left panel) and the Cauchy (on the right panel) are shown for different values of $\Theta$.}
    \label{horizons1}
\end{figure}

\begin{table}[!h]
\begin{center}
\caption{\label{evenntthorizon} The event horizon, $r_{+}$, is shown for a range of values of mass \(M\), and parameter \(\Theta\).}
\begin{tabular}{c c c ||| c c c c } 
 \hline\hline
 $M$ & $\Theta$ & $r_{+}$ & $M$ & $\Theta$ & $r_{+}$ &   \\ [0.2ex] 
 \hline 
 0.00 & 0.01 & 0.00000 & 2.00 & 0.00 & 4.00000 &  \\ 

 1.00 & 0.01 & 1.74071 & 2.00 & 0.01 & 3.75991 & \\
 
 2.00 & 0.01 & 3.75991 & 2.00 & 0.02 & 3.65027 &  \\
 
 3.00 & 0.01 & 5.76513 & 2.00 & 0.03 & 3.56092 &    \\
 
 4.00 & 0.01 & 7.76757 & 2.00 & 0.04 & 3.48142 &   \\
 
 5.00 & 0.01 & 9.76899 & 2.00 & 0.05 & 3.40766 &  \\
 
 6.00 & 0.01 & 11.7699 & 2.00 & 0.06 & 3.33747 &  \\
 
 7.00 & 0.01 & 13.7706 & 2.00 & 0.07 & 3.26952 &   \\
 
 8.00 & 0.01 & 15.7710 & 2.00 & 0.08 & 3.20282 &  \\
 
 9.00 & 0.01 & 17.7714 & 2.00 & 0.09 & 3.13661 &   \\
  10.0 & 0.01 & 19.7717 & 2.00 & 0.10 & 3.07023 &   \\
[0.2ex] 
 \hline \hline
\end{tabular}
\end{center}
\end{table}

\begin{table}[!h]
\begin{center}
\caption{\label{couchyhorizonzz} The Cauchy horizon, $r_{-}$, is shown for a range of values of mass \(M\), and parameter \(\Theta\).}
\begin{tabular}{c c c ||| c c c c } 
 \hline\hline
 $M$ & $\Theta$ & $r_{-}$ & $M$ & $\Theta$ & $r_{-}$ &   \\ [0.2ex] 
 \hline 
 0.00 & 0.01 & 0.000000 & 2.00 & 0.00 & 0.000000 &  \\ 

 1.00 & 0.01 & 0.259292 & 2.00 & 0.01 & 0.240086 & \\
 
 2.00 & 0.01 & 0.240086 & 2.00 & 0.02 & 0.349732 &  \\
 
 3.00 & 0.01 & 0.234870 & 2.00 & 0.03 & 0.439080 &    \\
 
 4.00 & 0.01 & 0.232429 & 2.00 & 0.04 & 0.518584 &   \\
 
 5.00 & 0.01 & 0.231013 & 2.00 & 0.05 & 0.592345 &  \\
 
 6.00 & 0.01 & 0.230088 & 2.00 & 0.06 & 0.662526 &  \\
 
 7.00 & 0.01 & 0.229436 & 2.00 & 0.07 & 0.730484 &   \\
 
 8.00 & 0.01 & 0.228952 & 2.00 & 0.08 & 0.797183 &  \\
 
 9.00 & 0.01 & 0.228579 & 2.00 & 0.09 & 0.863387 &   \\
  10.0 & 0.01 & 0.228281 & 2.00 & 0.10 & 0.929766 &   \\
[0.2ex] 
 \hline \hline
\end{tabular}
\end{center}
\end{table}

%%%%%%%%%%%%%%%%%%%%%%%%%%%%%%%%%%%%%%%%%%%%%%%%%%%%%%%%%%%%%%%%%%%%%%%%%%%%%%%%%%%%%%%%%%%%%%%%%%%%%%%%%%%%%%%%%%%%%%%%%%%%%%%%%%%%%%%%%%%%%%%%%%%%%%%%%%%%%%%%%%%%%%%%%%%%%%%%%%%%%%%%%%%%%%%%%%%%%%%%%%%%%%%%%%%%%%%%%%%%%%%%%%%%%%%%%%%%%%%%%%%%%%%%%%%%%%%%%%%%%%%%%%%%%%%%%%%%%%%%%%%%%%%%%%%%%%%%%%%%%%%%%%%%%%%%%%%%%%%%%%%%%%%%%%%%%%%%%%%%%%%%%%%%%%%%%%%%%%%%%%%%%%%%%%%%%%%%%%%%%%%%%%%%%%%%%%%%%%%%%%%%%%%%%%%%%%%%%%%%%%%%%%%%%%

\section{Thermodynamics, evaporation and emission rate}

In this section, we shall analyze the thermodynamic behavior of the system and the respective evaporation process; also, we shall estimate the lifetime of the black hole under consideration. 

%%%%%%%%%%%%%%%%%%%%%%%%%%%%%%%%%%%%%%%%%%%%%%%%%%%%%%%%%%%%%%%%%%%%%%%%%%%%%%%%%%%%%%%%%%%%%%%%%%%%%%%%%%%%%%%%%%%%%%%%%%%%%%%%%%%%%%%%%%%%%%%%%%%%%%%%%%%%%%%%%%%%%%%%%%%%%%%%

\subsection{Hawking temperature via topological method \label{tee}}

Using the topological technique, one can determine the \textit{Hawking} temperature without sacrificing the knowledge about the higher-dimensional space by using the Euclidean geometry of the 2-dimensional spacetime. The thermodynamic property of \textit{Hawking} temperature for a two-dimensional black hole can be established using the topological method \cite{Robson:2018con,Ovgun:2019ygw,Zhang:2020kaq}
\begin{equation}
T_{\text{H}}=\frac{\hbar c}{4\pi \chi k_{\text{B}}}\Sigma_{j\leq\chi}\int_{r_{h_j}}{\sqrt{|g|}\mathcal{R}\mathrm{d}r}.
\label{temperature}
\end{equation}
In this context, the symbols $\hbar$, $c$, and $k_{\text{B}}$ represent the Planck constant, speed of light, and Boltzmann's constant, respectively. Additionally, $g$ corresponds to the metric determinant, and $r_{h_j}$ signifies the $j$-th Killing horizon. For this study, we adopt the values $\hbar=1$, $c=1$, and $k_{\text{B}}=1$ for these parameters. The function $\mathcal{R}$ denotes the Ricci scalar in the two-dimensional spacetime. The variable $\chi$ represents the Euler characteristic of the Euclidean geometry and is linked to the count of Killing horizons. The symbol $\Sigma_{j\leq\chi}$ signifies the summation across the Killing horizons.

The Euler characteristic in a two-dimensional spacetime is expressed as follows:

\begin{equation}
\chi=\int \sqrt{|g|} \mathrm{d}^2x\frac{\mathcal{R}}{4\pi}.
\end{equation}

Upon employing the Wick rotation $t=i\tau$ and defining the new compact time as the inverse temperature $\beta$, the Euler characteristic $\chi$ takes the form \cite{Robson:2018con,Ovgun:2019ygw}.
\begin{equation}
\chi=\int_0^\beta{\mathrm{d}\tau}\int_{r_{\text{H}}}\sqrt{|g|}\mathrm{d}r \frac{\mathcal{R}}{4\pi}.
\end{equation}

Subsequently, the connection between the \textit{Hawking} temperature $T_{H}$ and the Euler characteristic $\chi$ is established through the relation:

\begin{equation}
\frac{1}{4\pi T_{H}}\int_{r_{\text{H}}}{\sqrt{|g|}\mathcal{R}\mathrm{d}r}=\chi,
\end{equation}
this relationship serves as the basis for Eq. (\ref{temperature}). 

By considering a specific hypersurface, the black hole can be transformed into a two-dimensional configuration with a reduced metric \cite{Achucarro:1993fd} through the Wick rotation ($\tau = i t$):
\begin{equation}
\mathrm{d}s^2=f_{\Theta} (r) \mathrm{d}\tau^2+\frac{\mathrm{d}r^2}{f_{\Theta} (r) }.
\label{weakmetric}
\end{equation}

The Ricci scalar corresponding to the reduced metric (\ref{weakmetric}) is given by:
\begin{equation}
\mathcal{R}=\frac{4 M}{r^3}-\frac{48 \sqrt{\Theta } M}{\sqrt{\pi } r^4}.
\end{equation}

Hence, the temperature of the black hole is determined by employing the formula:
\begin{equation}
\ {T_\Theta }=\frac{1}{4 \pi \chi} \int_{r_{+}} \sqrt{|g|} R \mathrm{d} r= \frac{M}{2 \pi r_{+}^2}-\frac{4 \sqrt{\Theta } M}{\pi^{3/2}r_{+}^3},
\end{equation}
which leads to the same result by using the surface gravity formula
\ie
\label{hawkingtemperature}
\ {T_\Theta } = \frac{1}{{4\pi \sqrt {{g_{00}}{g_{11}}} }}{\left. {\frac{{\mathrm{d}{g_{00}}}}{{\mathrm{d}r}}} \right|_{r = {r_{+ }}}} = \frac{M}{2 \pi r_{+}^2}-\frac{4 \sqrt{\Theta } M}{\pi ^{3/2}r_{+}^3},
\fe
where $M$ is obtained if we consider $f(r_{+})=0$, which reads
\ie
M = \frac{r_{+}^2}{2 \left(r_{+}-\frac{4 \sqrt{\Theta }}{\sqrt{\pi }}\right)}.
\fe
Explicitly, in terms of the horizon, such a thermal quantity is given by
\ie
T_{\Theta} = \frac{1}{2 \pi  r_{+} \left(\frac{r_{+}}{r_{+}-\frac{8 \sqrt{\Theta }}{\sqrt{\pi }}}+1\right)}.
\fe

\begin{figure}
    \centering
     \includegraphics[scale=0.385]{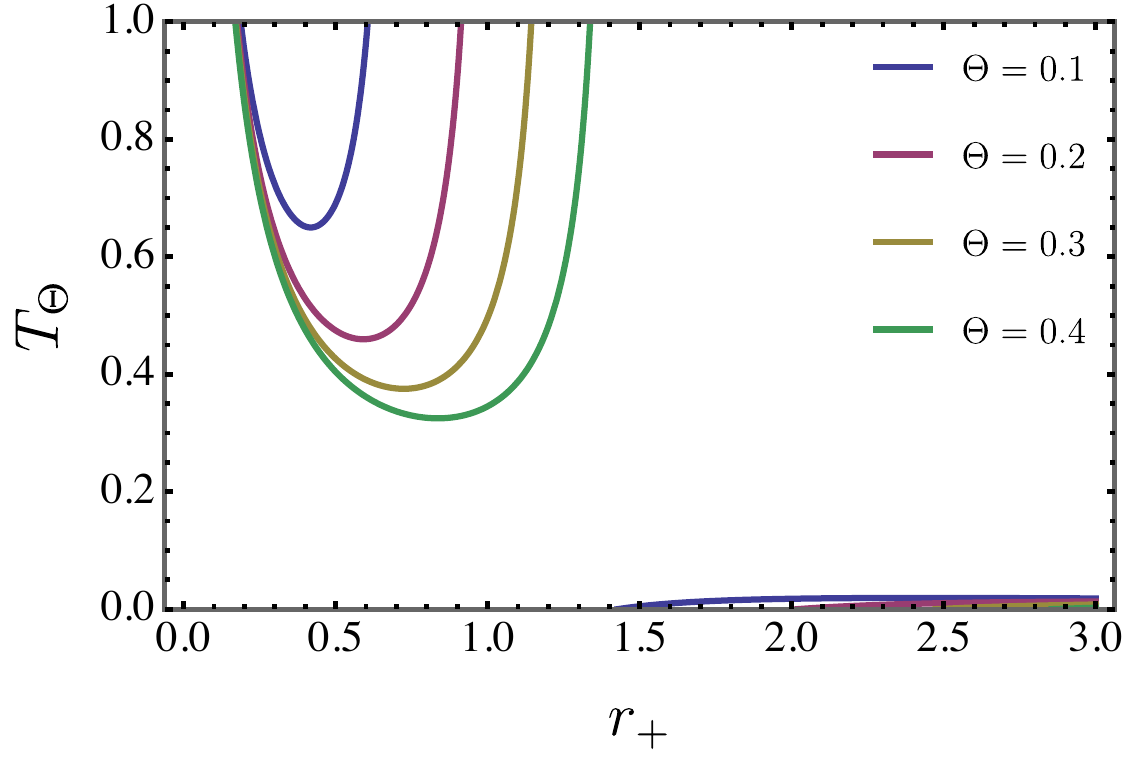}
     \includegraphics[scale=0.4]{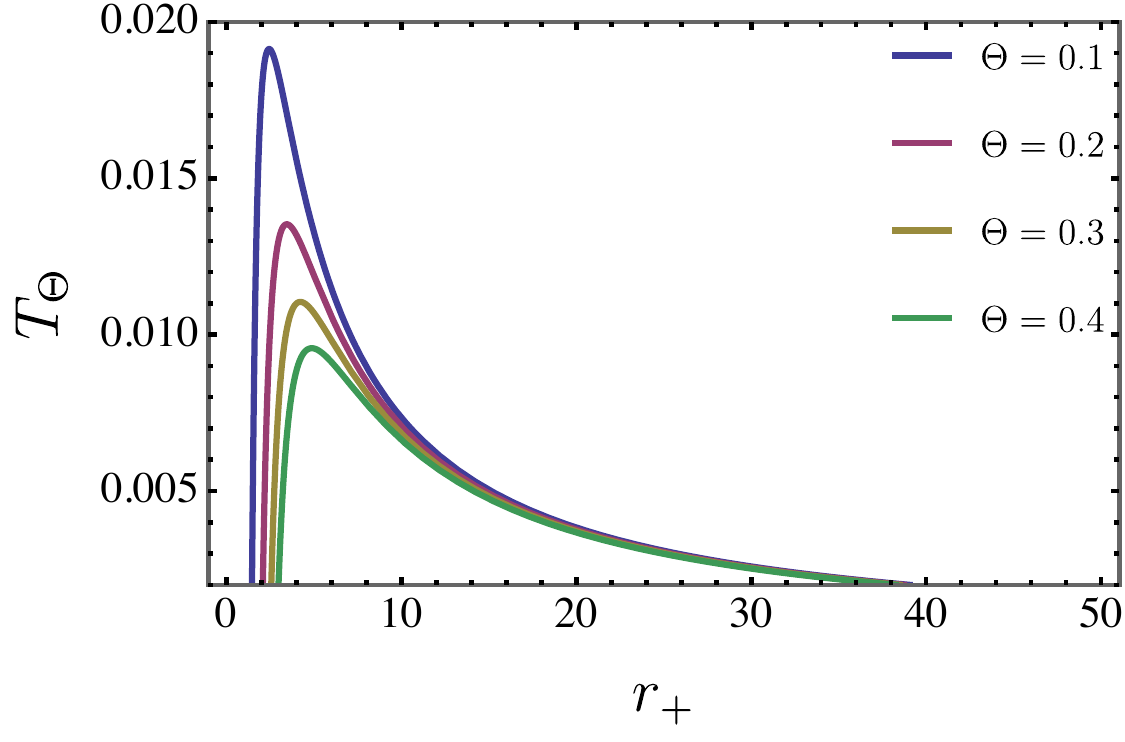}
    \includegraphics[scale=0.4]{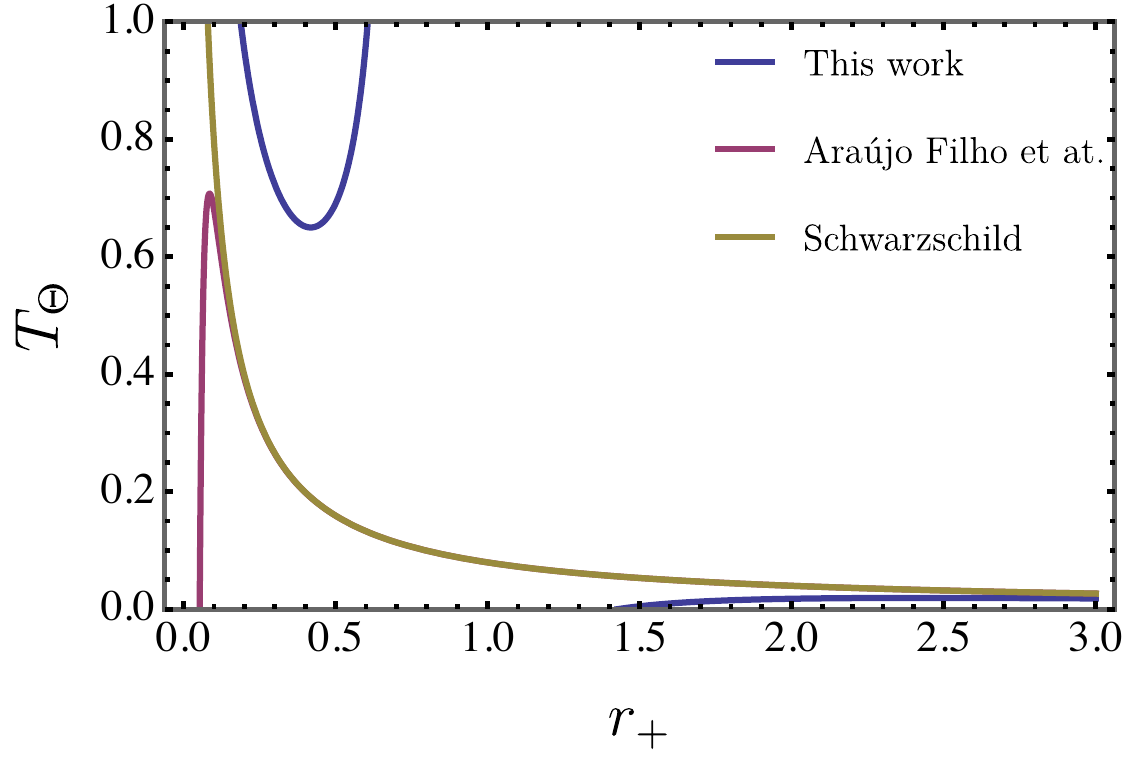}
    \caption{The \textit{Hawking} temperature are shown for different ranges of $r_{+}$ and $\Theta$.}
    \label{hawkingtemperatures}
\end{figure}

To visualize better the \textit{Hawking} temperature, we present Fig. \ref{hawkingtemperatures}, regarding different ranges of the event horizon and the non-commutative parameter. Notably, note that, for small values of $r_{+}$, such a thermodynamic function indicates a phase transition, as displayed in the left panel. In general lines, $\Theta$ decreases as the magnitude of $T_{\Theta}$ grows, as seen in the right panel. Furthermore, we compare our results with those of the Schwarzschild case and another recent study conducted within the framework of non-commutative gauge theory \cite{t29}. In addition, we can also write $T_{\Theta}$ as function of the mass, as follows
\ie
T_{\Theta} = \frac{1}{4 \pi ^{3/4} M \left(\frac{\sqrt{\pi } M-4 \sqrt{\Theta }}{\sqrt{M \left(\sqrt{\pi } M-8 \sqrt{\Theta }\right)}}+\sqrt[4]{\pi }\right)}.
\fe
In this context, we can directly compare the expression for the \textit{Hawking} temperature obtained in Ref. \cite{t29} with our results more naturally. This feature is displayed in Tab. \ref{thermo}.

%%%%%%%%%%%%%%%%%%%%%%%%%%%%%%%%%%%%%%%%%%%%%%%%%%%%%%%%%%%%%%%%%%%%%%%%%%%%%%%%%%%%%%%%%%%%%%%%%%%%%%%%%%%%%%%%%%%%%%%%%%%%%%%%%%%%%%%%%%%%%%%%%%%%%%%%%%%%%%%%%%%%%%%%%%%%%%%%%%%%%%%%%%%%%%%%%%%%%%%%%%%%%%%%%%%%%%%%%%%%%%%%%%%%%%%%%%

\subsection{Entropy}

\begin{figure}
    \centering
     \includegraphics[scale=0.39]{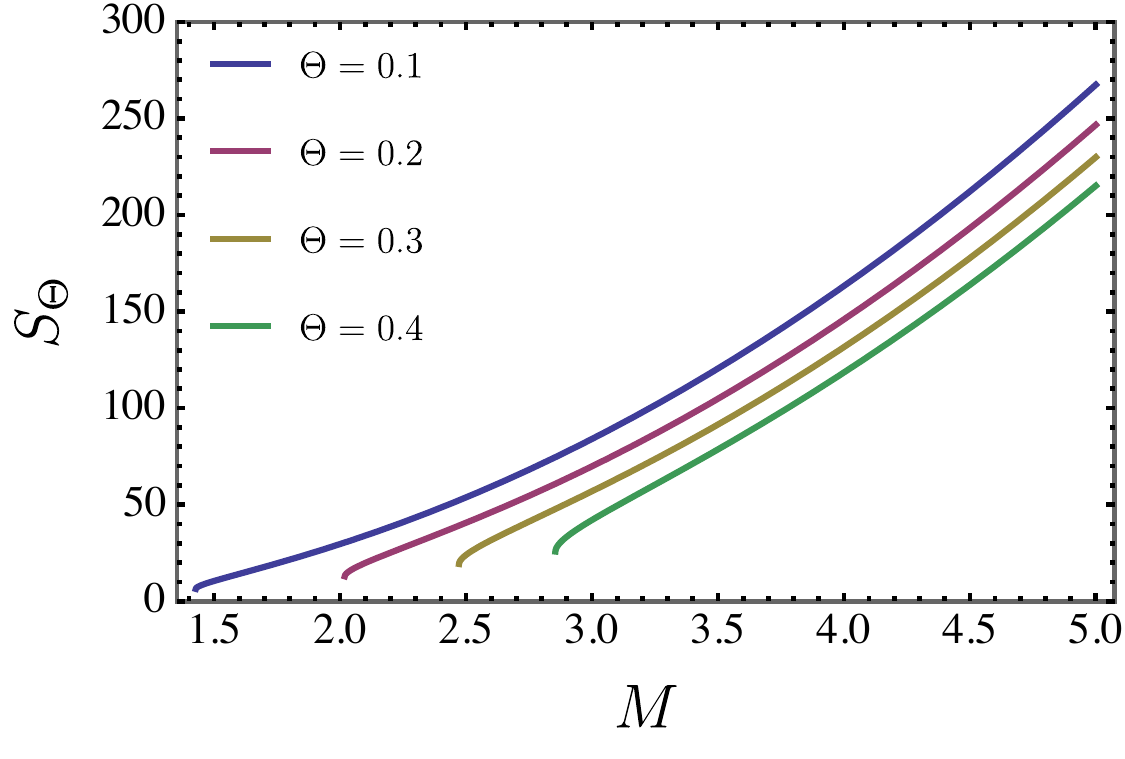}
     \includegraphics[scale=0.4]{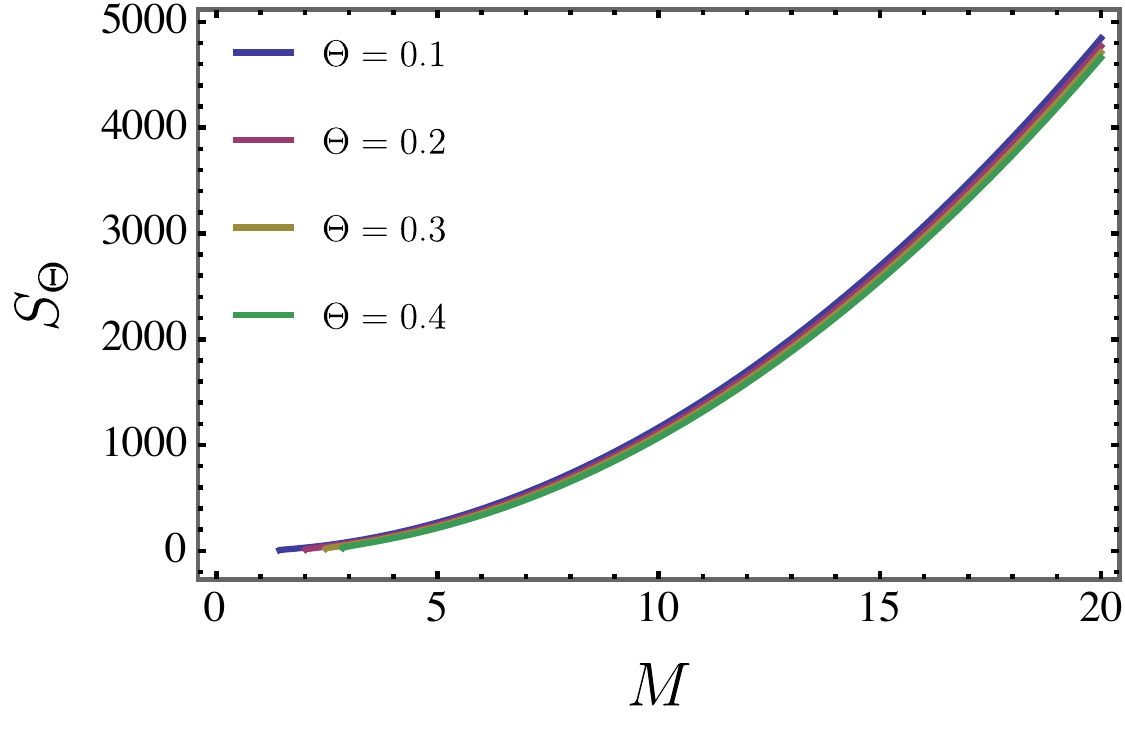}
     \includegraphics[scale=0.4]{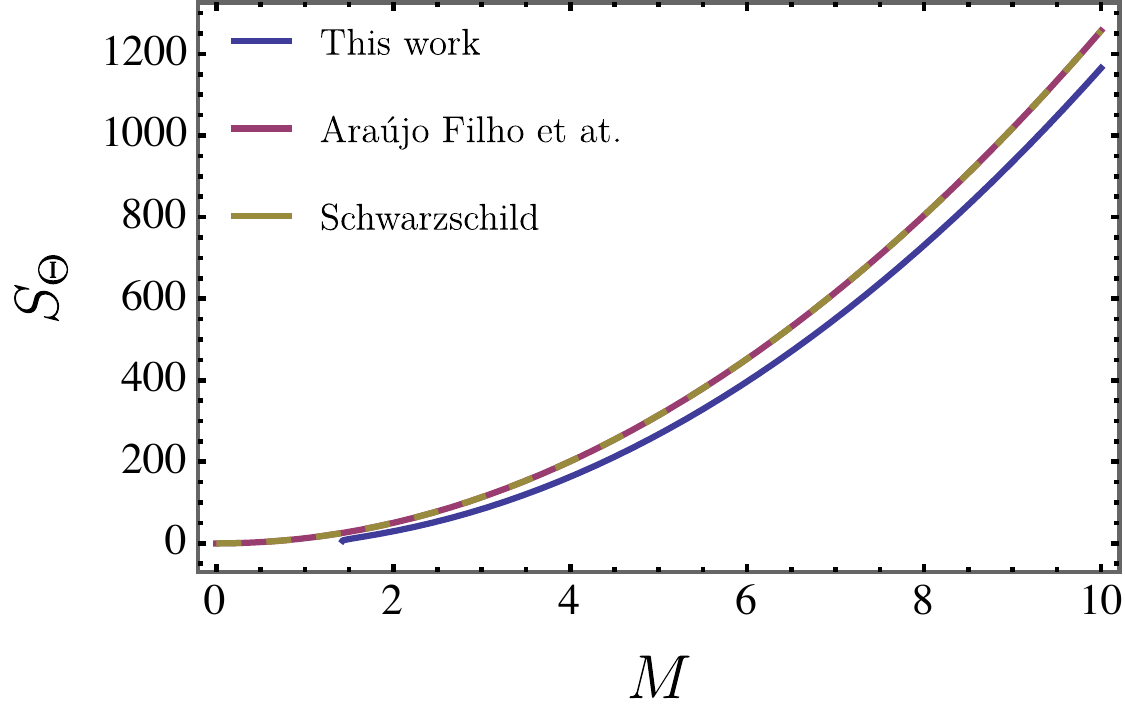}
    \caption{The entropy is exhibited for different ranges of $M$ and $\Theta$.}
    \label{entropies}
\end{figure}

Another thermodynamic quantity worthy to be investigated is indeed the entropy. Therefore, the area of the event horizon can be cast as:
\ie
\ {A_\Theta } =\int {\int {\sqrt {{g_{22}}{g_{33}}} } } \mathrm{d}\theta \mathrm{d}\varphi = 4\pi r_{+}^2.
\fe
From this result, we can properly compute the entropy, as shown below
\ie
\label{entropy}
\ {S_\Theta } = \frac{{{A_\Theta }}}{4} = \pi  \left(\frac{\sqrt{M \left(\sqrt{\pi } M-8 \sqrt{\Theta }\right)}}{\sqrt[4]{\pi }}+M\right)^2.
\fe
In Fig. \ref{entropies}, we show the entropy behavior, considering distinct values of the mass and parameter $\Theta$. As we can see, the second law of thermodynamics is also verified for our black hole solution. Furthermore, we compare our outcomes with the Schwarzschild case and Ref. \cite{t29}, which is shown in Tab. \ref{thermo}.

%%%%%%%%%%%%%%%%%%%%%%%%%%%%%%%%%%%%%%%%%%%%%%%%%%%%%%%%%%%%%%%%%%%%%%%%%%%%%%%%%%%%%%%%%%%%%%%%%%%%%%%%%%%%%%%%%%%%%%%%%%%%%%%%%%%%%%%%%%%%%%%%%%%%%%%%%%%%%%%%%%%%%%%%%%%%%%%%%%%%%%%%%%%%%%%%%%%%%%%%%%%%%%%%%%%%%%%%%%%%%%%%%%%%%%%%%%

\subsection{Heat capacity}

\begin{figure}
    \centering
     \includegraphics[scale=0.4]{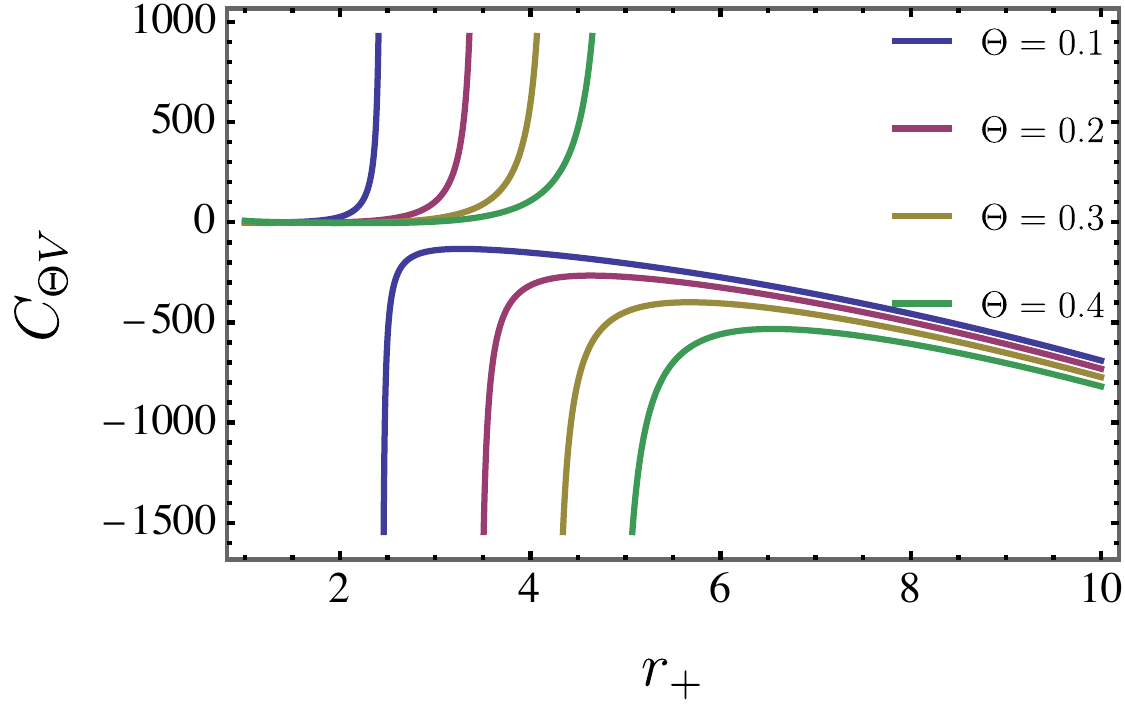}
     \includegraphics[scale=0.41]{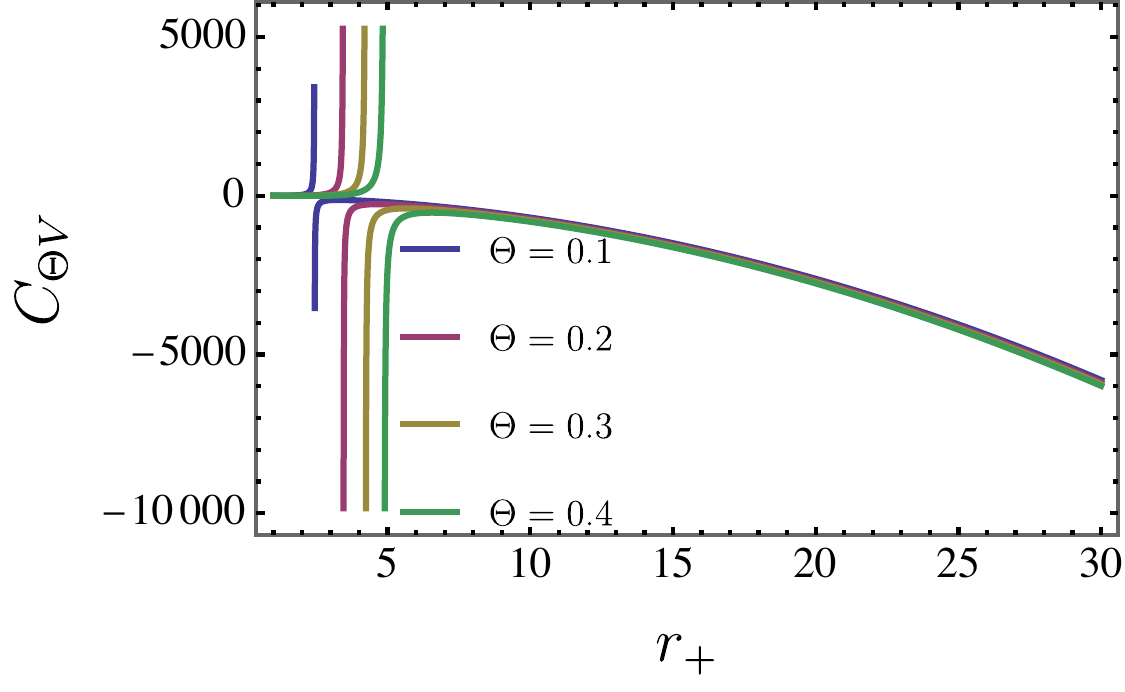}
     \includegraphics[scale=0.41]{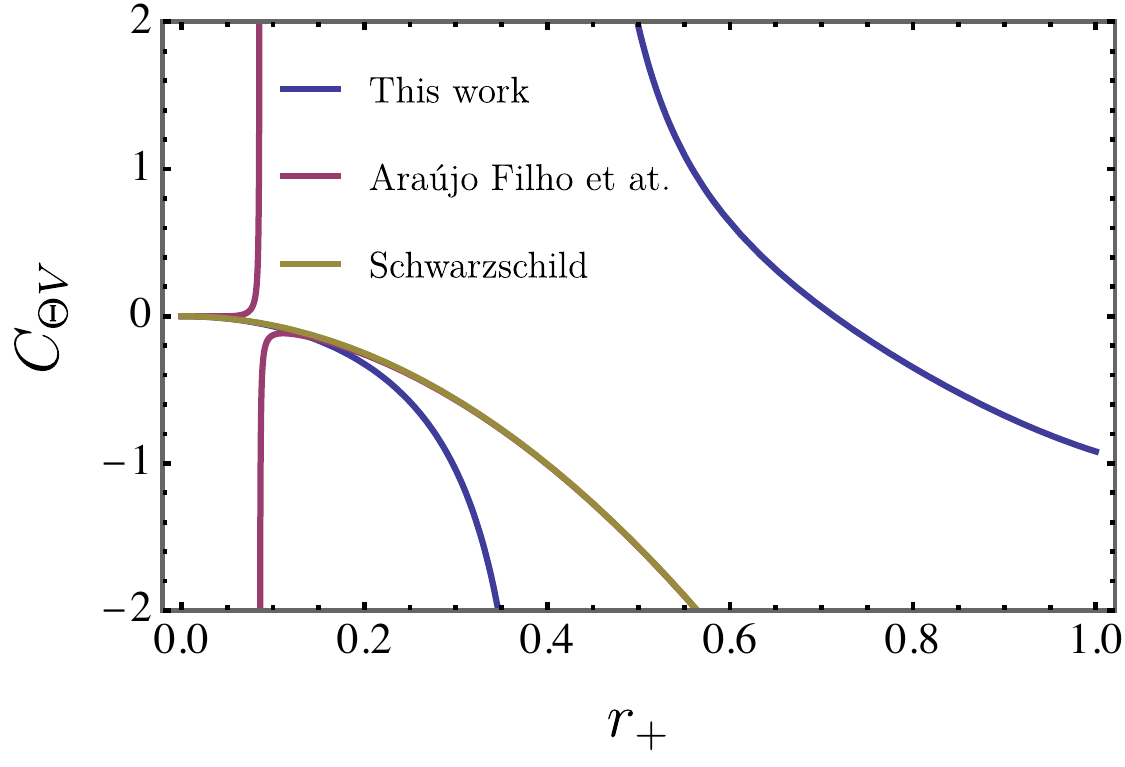}
    \caption{The heat capacity is displayed for different ranges of $M$ and $\Theta$.}
    \label{heats}
\end{figure}

Finally, let us address the heat capacity
\ie
C_{\Theta V} = {T_\Theta }\frac{{\partial {S_\Theta }}}{{\partial {T_\Theta }}} = T_{\Theta} \frac{\partial S_{\Theta}/\partial M}{\partial T_{\Theta}/\partial M} = 2 \pi  r_{+}^2 \left(\frac{1}{4-\frac{32 \Theta +\pi  r_{+}^2}{4 \sqrt{\pi } \sqrt{\Theta } r_{+}}}-1\right).
\fe
In Fig. \ref{heats}, the behavior of the heat capacity analogously to what we have accomplished up to now is exhibited. Notice that several phase transitions are also highlighted in such a plot.  Finally, to display a general panorama of all thermodynamic functions, we present Tab. \ref{thermo}. Here, we compare all thermal quantities developed in this work with Schwarzschild and the results shown in Ref. \cite{t29}. For the sake of providing a direct comparison with such a reference and our outcomes, we write the heat capacity $C_{V}$ as a function of mass $M$, as follows
\ie
C_{V} = 2 \pi  \left(\frac{1}{-\frac{\sqrt{\pi } M}{4 \sqrt{\Theta }}+\frac{\sqrt{M \left(\sqrt{\pi } M-8 \sqrt{\Theta }\right)}}{\sqrt[4]{\pi } M}-\frac{\sqrt[4]{\pi } \sqrt{M \left(\sqrt{\pi } M-8 \sqrt{\Theta }\right)}}{4 \sqrt{\Theta }}+3}-1\right) \left(\frac{\sqrt{M \left(\sqrt{\pi } M-8 \sqrt{\Theta }\right)}}{\sqrt[4]{\pi }}+M\right)^2.
\fe

\begin{table}[t]
	\centering
		\caption{\label{thermo}Comparison of the thermodynamic properties between a NC Schwarzschild BH obtained by the present study (via deformed mass) and the previous results (from deformed metric). Here, $\Gamma = -\frac{\sqrt{\pi } M}{4 \sqrt{\Theta }}+\frac{\sqrt{M \left(\sqrt{\pi } M-8 \sqrt{\Theta }\right)}}{\sqrt[4]{\pi } M}-\frac{\sqrt[4]{\pi } \sqrt{M \left(\sqrt{\pi } M-8 \sqrt{\Theta }\right)}}{4 \sqrt{\Theta }}+3$.}
	\setlength{\arrayrulewidth}{0.3mm}
	\setlength{\tabcolsep}{10pt}
	\renewcommand{\arraystretch}{1}
	\begin{tabular}{c c c c}
		\hline \hline
		~ & This work & Araújo Filho et al. \cite{t29} & Schwarzschild \\ \hline
		$T_\Theta$ &  $1/\left[4 \pi ^{3/4} M \left(\frac{\sqrt{\pi } M-4 \sqrt{\Theta }}{\sqrt{M \left(\sqrt{\pi } M-8 \sqrt{\Theta }\right)}}+\sqrt[4]{\pi }\right) \right]$ & $ \frac{1}{8\pi M} - \frac{3}{512 \pi M^{3}} \Theta^{2}$ & $\frac{1}{4 \pi (2M)} $ \\
		$A_\Theta$ & $4\pi \left(\frac{\sqrt{\pi  M^2-8 \sqrt{\pi } \sqrt{\Theta } M}}{\sqrt{\pi }}+M\right)^{2}$ &$ 4\pi (2M)^{2} + \frac{5 \pi}{16} \Theta^{2}$  & $4 \pi (2M)^{2}$ \\ 
		$S_\Theta$ &  $\pi \left(\frac{\sqrt{\pi  M^2-8 \sqrt{\pi } \sqrt{\Theta } M}}{\sqrt{\pi }}+M\right)^{2}$  & $  \pi (2M)^{2} + \frac{5\pi}{64}\Theta^{2}  $ & $\pi (2M)^{2}$ \\ 
		$C_{V\Theta}$ &  $2  \pi  \left(\frac{1}{\Gamma}-1\right) \left(\frac{\sqrt{M \left(\sqrt{\pi } M-8 \sqrt{\Theta }\right)}}{\sqrt[4]{\pi }}+M\right)^2$ & $   - \frac{8M^{2} \pi (64M^{2} - 3\Theta^{2}) }{64M^{2} - 9 \Theta^{2}}   $ & $-2 \pi (2M)^{2}$ \\ \hline\hline
	\end{tabular}
\end{table}

%%%%%%%%%%%%%%%%%%%%%%%%%%%%%%%%%%%%%%%%%%%%%%%%%%%%%%%%%%%%%%%%%%%%%%%%%%%%%%%%%%%%%%%%%%%%%%%%%%%%%%%%%%%%%%%%%%%%%%%%%%%%%%%%%%%%%%%%%%%%%%%%%%%%%%%%%%%%%%%%%%%%%%%%%%%%%%%%%%%%%%%%%%%%%%%%%%%%%%%%%%%%%%%%%%%%%%%%%%%%%%%%%%%%%%%%%%%%%%%%%%%%%%%%%%%%%%%%%%%%%%%%%%%%%%%%%%%%%%%%%%%%%%%%%%%%%%%%%%%%%%%%%%%%%%%%%%%%%%%%%%%%%%%%%%%%%%%%%%%%%%%%%%%%%%%%%%%%%%%%%%%%%%%%%%%%%%%%%%%%%%%%%%%%%%%%%%%%%%%%%%%%%%%%%%%%%%%%%%%%%%%%%%%%%%

\subsection{Evaporation process}

\begin{figure}
    \centering
      \includegraphics[scale=0.45]{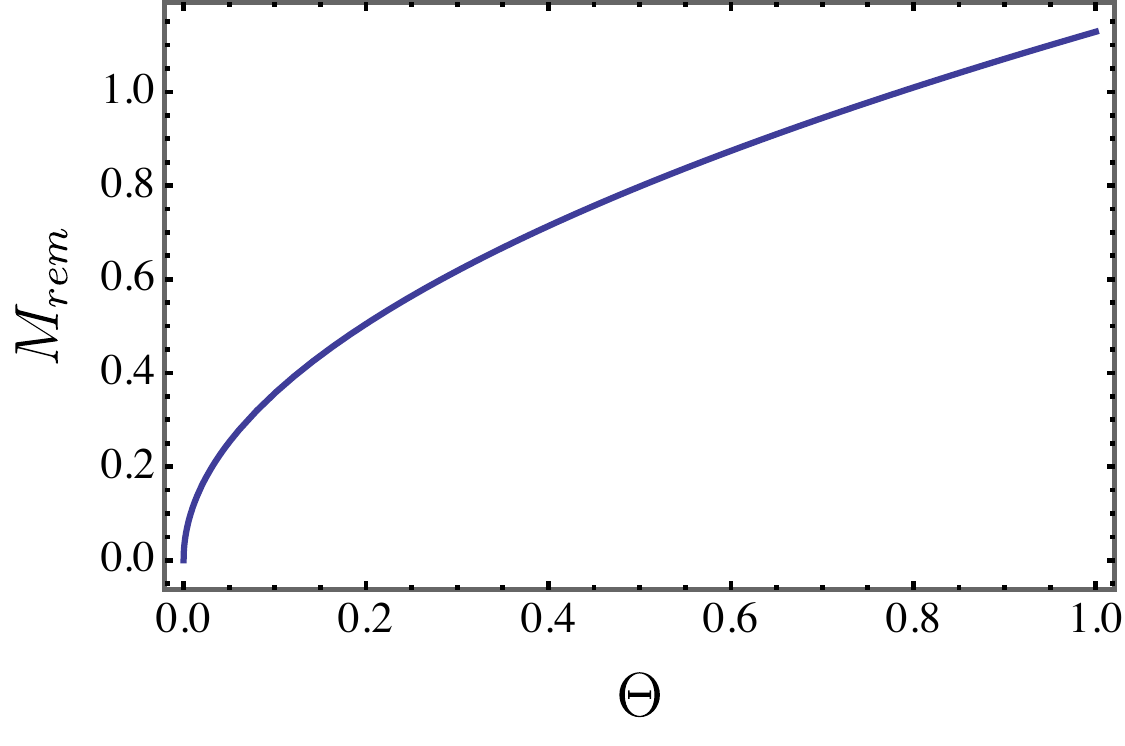}
    \caption{$M_{rem}$ as a function of the non-commutative parameter $\Theta$.}
    \label{ramananttttt}
\end{figure}

Now, let us examine the evaporation process of our non-commutative black hole. To do so, it is reasonable to calculate the remnant mass, $M_{rem}$. Note that, in our scenario, as the black hole approaches its final stage of evaporation ($T_{\Theta} \to 0$), this culminates in the following expression to the mass
\ie
\label{regnnmavnatsmass}
M_{rem} = \frac{2 \sqrt{\Theta }}{\sqrt{\pi }}.
\fe
From the aforementioned equation, it is clear that only one parameter, namely $\Theta$, exerts influence on the modification of $M_{\text{rem}}$. To facilitate a comprehensive understanding of the behavior of this mass, we present Fig. \ref{ramananttttt}.
Another important aspect worth exploring in this context is essentially the black hole lifetime. Thereby, we have
\ie
\label{evava}
\frac{\mathrm{d}M}{\mathrm{d}\tau} = - \alpha \sigma a T_{\Theta}^{4}.
\fe
Here, $a$ represents the radiation constant, $\sigma$ denotes the cross-sectional area, and $\alpha$ is for the \textit{grey-body} factor. Taking into consideration the geometric optics approximation, $\sigma$ is thereby understood as the cross section for photon capture:
\ie
\sigma = \pi \left. \left( \frac{g_{\varphi\varphi}}{g_{tt}} \right)  \right|_{r = {r_{ph}}}= \frac{\left(\sqrt{M \left(9 \sqrt{\pi } M-64 \sqrt{\Theta }\right)}+3 \sqrt[4]{\pi } M\right)^4}{8 \left(3 \pi  M^2-16 \sqrt{\pi } \sqrt{\Theta } M+\pi ^{3/4} M \sqrt{M \left(9 \sqrt{\pi } M-64 \sqrt{\Theta }\right)}\right)},
\fe
where $r_{ph}$ is the photon sphere radius given by
\ie
r_{ph} = \frac{\left(\sqrt{M \left(9 \sqrt{\pi } M-64 \sqrt{\Theta }\right)}+3 \sqrt[4]{\pi } M\right)^4}{8 \left(3 \pi  M^2-16 \sqrt{\pi } \sqrt{\Theta } M+\pi ^{3/4} M \sqrt{M \left(9 \sqrt{\pi } M-64 \sqrt{\Theta }\right)}\right)}.
\fe
With this, we can substitute these previous expressions in Eq. (\ref{evava}), so that
\ie
\begin{split}
\frac{\mathrm{d}M}{\mathrm{d}\tau} & = 
\frac{-\xi M \left(\sqrt{\pi } M-8 \sqrt{\Theta }\right)^2}{8 \pi ^{7/2} \left(\sqrt{M \left(\sqrt{\pi } M-8 \sqrt{\Theta }\right)}+\sqrt[4]{\pi } M\right)^4 \left(-16 \sqrt{\Theta }+\sqrt[4]{\pi } \sqrt{M \left(9 \sqrt{\pi } M-64 \sqrt{\Theta }\right)}+3 \sqrt{\pi } M\right)},
\end{split}
\fe
where $\xi = a \alpha$. Therefore, we have to solve
\ie
\begin{split}
& \int_{0}^{t_{\text{evap}}} \xi \mathrm{d}\tau  = \\
& \int_{M_{i}}^{M_{rem}}
\mathrm{d}M\left[ \frac{-\xi M \left(\sqrt{\pi } M-8 \sqrt{\Theta }\right)^2}{8 \pi ^{7/2} \left(\sqrt{M \left(\sqrt{\pi } M-8 \sqrt{\Theta }\right)}+\sqrt[4]{\pi } M\right)^4 \kappa}\right]^{-1} ,
\end{split}
\fe
where $M_{i}$ is the initial mass configuration, $t_{\text{evap}}$ being the time for reaching the final stage of the respective its evaporation, and $\kappa = \left(-16 \sqrt{\Theta }+\sqrt[4]{\pi } \sqrt{M \left(9 \sqrt{\pi } M-64 \sqrt{\Theta }\right)}+3 \sqrt{\pi } M\right)$. Therefore, we obtain the general expression for $t_{\text{evap}}$, as shown in the Appendix. Now, let us estimate the lifetime of the black hole under consideration for a particular configuration of $M_{i}$ (initial mass) and $M_{f}$ (final mass). For instance, if $M_{i} = 10$ and $M_{f}=M_{rem} = 0.356825$ (for $\Theta = 0.001$), we get
\ie
t_{evap} = \frac{1}{\xi} \left( 2.51807\times 10^7 \right).
\fe

\begin{figure}
    \centering
      \includegraphics[scale=0.42]{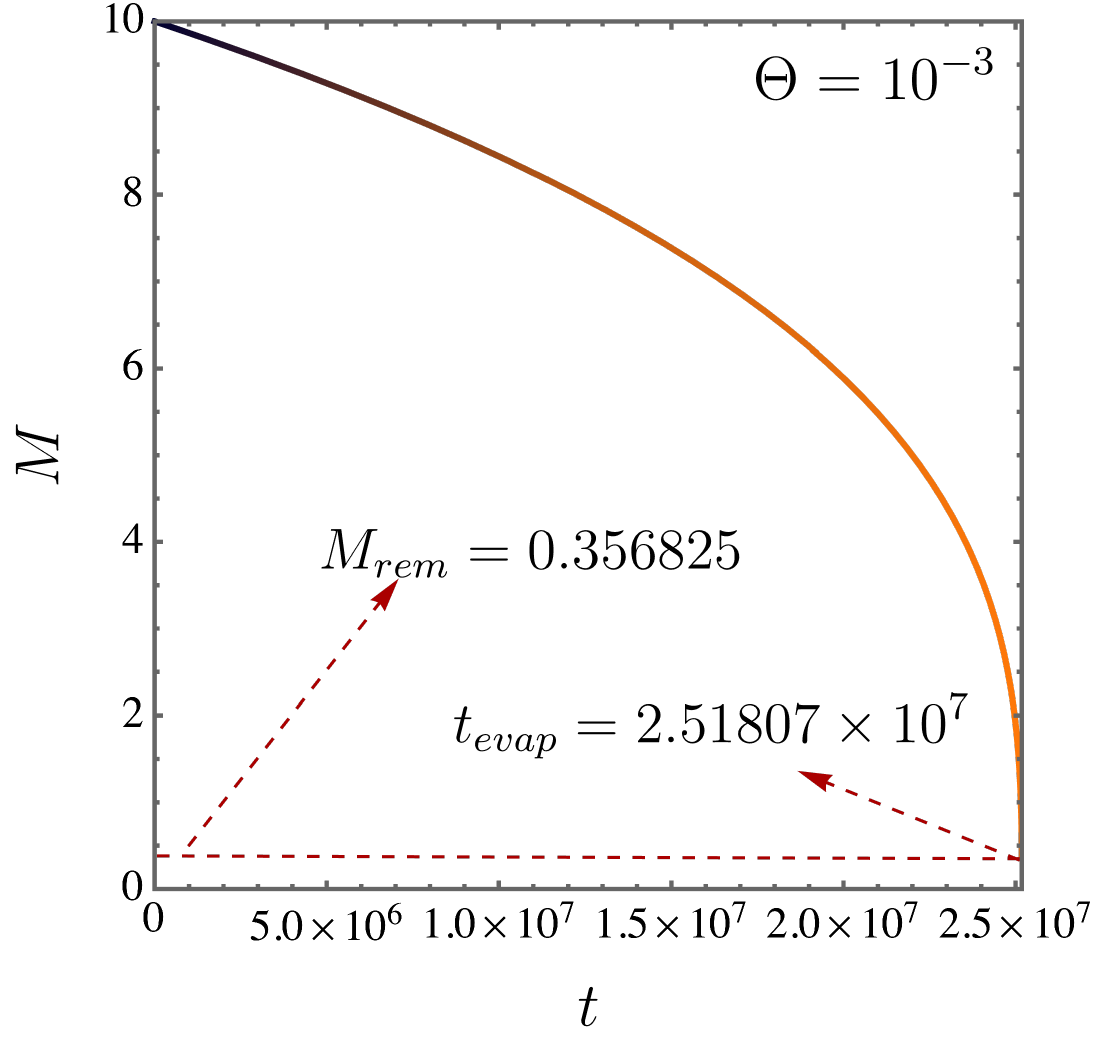}
    \caption{The reduction of mass as a function of time. }
    \label{reductionofmass}
\end{figure}

In Fig. \ref{reductionofmass}, we show the reduction of mass $M$ until reaching the final state of the black hole evaporation, i.e., when it reaches its remnant mass $M_{rem}$.

\begin{figure}
    \centering
      \includegraphics[scale=0.42]{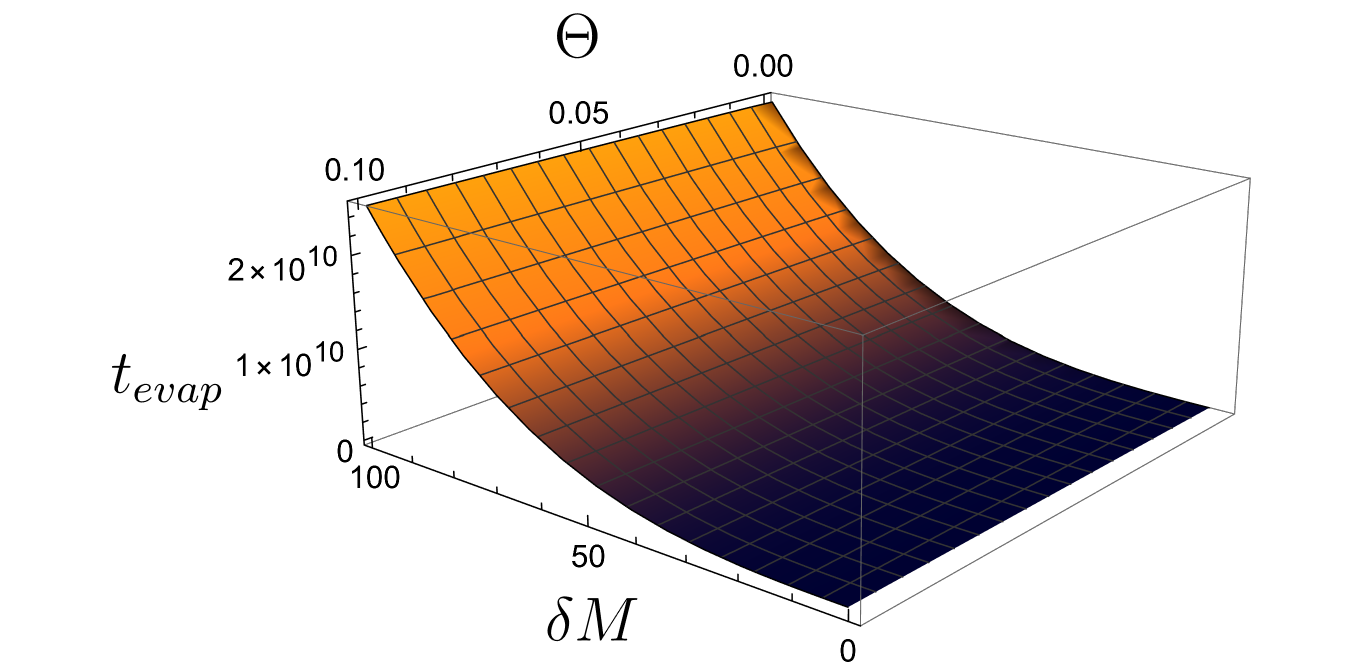}
       \includegraphics[scale=0.35]{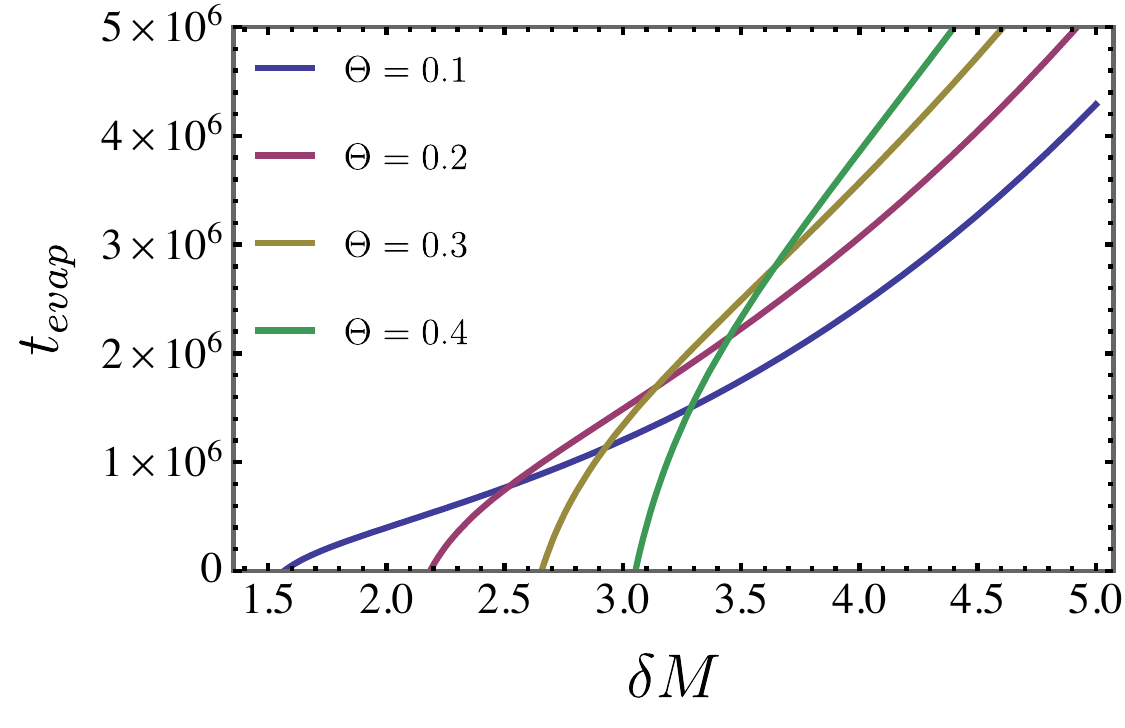}
    \caption{The evaporation time, denoted as $t_{evap}$, as a function of $M$ and $\Theta$ on the left hand; also, it is also displayed $t_{evap}$, regarding some fixed values of $\Theta$ as a function of $M$ .}
    \label{ramananttttt1}
\end{figure}

%%%%%%%%%%%%%%%%%%%%%%%%%%%%%%%%%%%%%%%%%%%%%%%%%%%%%%%%%%%%%%%%%%%%%%%%%%%%%%%%%%%%%%%%%%%%%%%%%%%%%%%%%%%%%%%%%%%%%%%%%%%%%%%%%%%%%%%%%%%%%%%%%%%%%%%%%%%%%%%%%%%%%%%%%%%%%%%%%%%%%%%%%%%%%%%%%%%%%%%%%%%%%%%%%%%%%%%%%%%%%%%%%%%%%%%%%%%%%%%%%%%%%%%%%%%%%%%%%%%%%%%%%%%%%%%%%%%%%%%%%%%%%%%%%%%%%%%%%%%%%%%%%%%%%%%%%%%%%%%%%%%%%%%%%%%%%%%%%%%%%%%%%%%%%%%%%%%%%%%%%%%%%%%%%%%%%%%%%%%%%%%%

\subsection{Emission rate}

Subsequently, our focus shifts to determining the rate of energy emission by black holes, commonly referred to as \textit{Hawking} radiation. It is established that the black hole shadow corresponds to its high energy absorption cross-section for observers at infinity. Conversely, at higher energies, it converges to a constant value known as $\sigma_{lim}$, which, for a spherically symmetric black hole, is described in \cite{wei2013observing}.
\begin{equation}
	\sigma_{lim} \approx R_{sh}^{2}.
\end{equation}

Applying such a limit, the computation of the rate of energy emission of a black hole is then performed\cite{wei2013observing,papnoi2022rotating,panah2020charged,sau2023shadow} 
\begin{equation}\label{emission}
	\frac{{{\mathrm{d}^2}E}}{{\mathrm{d}\omega \mathrm{d}t}} = \frac{{2{\pi ^2}\sigma_{lim} }}{{e^{{\omega }/T_\Theta} - 1}}{\omega ^3},
\end{equation}
with $\omega$ represents the photon frequency.

\begin{figure}[h!]
    \centering
    \includegraphics[scale=0.418]{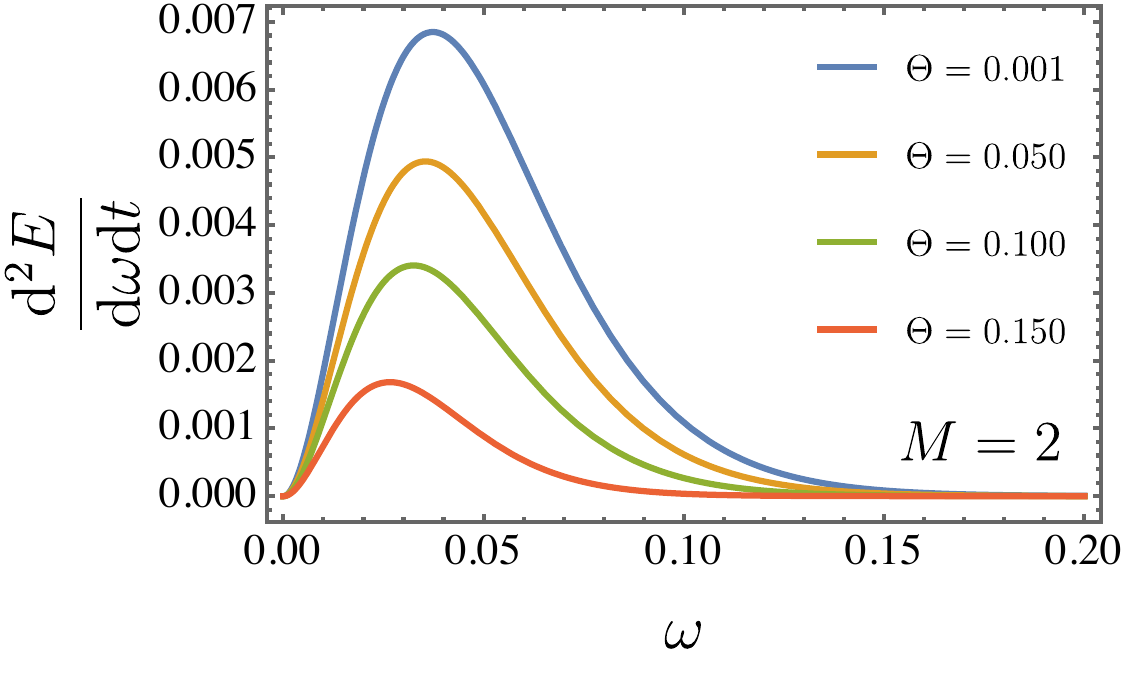}
    \includegraphics[scale=0.425]{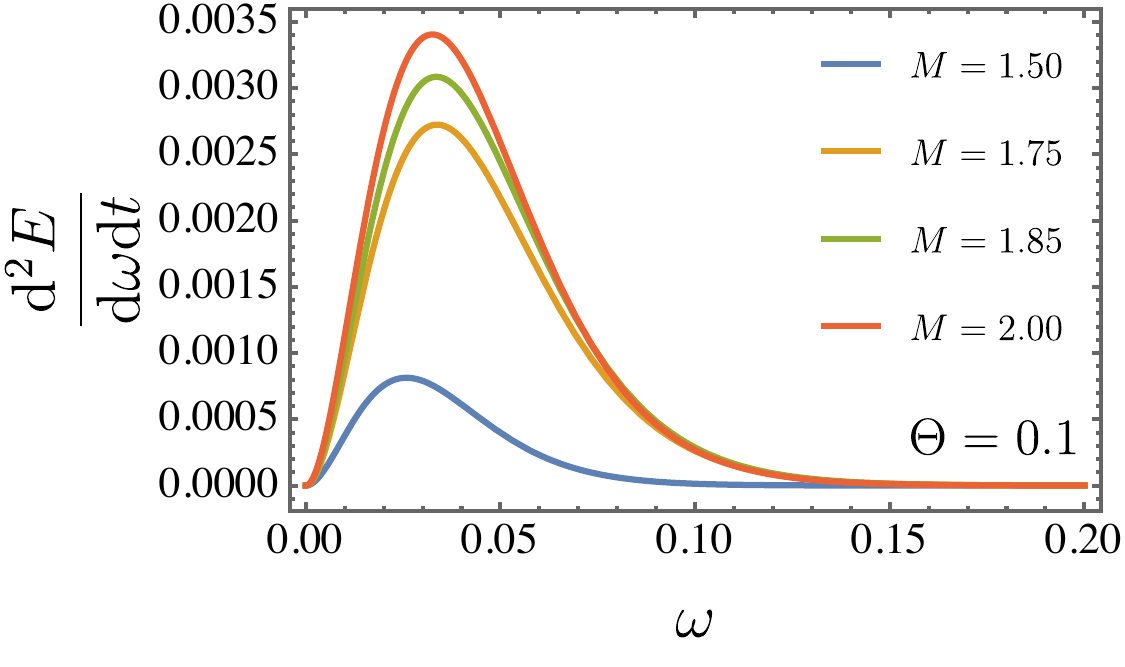}
    \caption{The emission rate for distinct patterns of $\Theta$ and $M$.}
    \label{fig:emissionrate}
\end{figure}

Figure \ref{fig:emissionrate} illustrates the emission rate plotted against the frequency $\omega$ for different values of $\Theta$. It is evident that as frequencies tend towards both zero and infinity, the emission rate diminishes. Additionally, it is noteworthy that $\Theta$ contributes to diminishing the magnitude of the emission rate.

%%%%%%%%%%%%%%%%%%%%%%%%%%%%%%%%%%%%%%%%%%%%%%%%%%%%%%%%%%%%%%%%%%%%%%%%%%%%%%%%%%%%%%%%%%%%%%%%%%%%%%%%%%%%%%%%%%%%%%%%%%%%%%%%%%%%%%%%%%%%%%%%%%%%%%%%%%%%%%%%%%%%%%%%%%%%%%%%%%%%%%%%%%%%%%%%%%%%%%%%%%%%%%%%%%%%%%%%%%%%%%%%%%%%%%%%%%%%%%%%%%%%%%%%%%%%%%%%%%%%%%%%%%%%%%%%%%%%%%%%%%%%%%%%%%%%%%%%%%%%%%%%%%%%%%%%%%%%%%%%%%%%%%%%%%%%%%%%%%%%%%%%%%%%%%%%%%%%%%%%%%%%%%%%%%%%%%%%%%%%%%%%%%%%%%%%%%%%%%%%%%%%%%%%%%%%%%%%%%%%%%%%%%%%%%

\section{Quasinormal modes \label{quasi}}

A noteworthy phenomenon termed \textit{quasinormal} modes emerges, revealing distinct oscillation patterns that remain largely unaltered by initial perturbations. These modes exemplify the inherent characteristics of the system, deriving from the natural oscillations of spacetime, irrespective of specific initial conditions. Unlike \textit{normal} modes associated with closed systems, \textit{quasinormal} modes correspond to open systems, gradually dissipating energy through the emission of gravitational waves. Mathematically, they are characterized as poles of the complex Green function.

To ascertain their frequencies, researchers derive solutions to the wave equation within a system governed by the background metric \(g_{\mu\nu}\). However, obtaining analytical solutions for these modes frequently poses significant challenges. Various methodologies have been explored in scientific literature to address this issue. Among these, the WKB (Wentzel-Kramers-Brillouin) method stands out as particularly prevalent. Its origins can be traced back to the seminal contributions of Will and Iyer \cite{Iyer:1986np,Iyer:1986nq}. Subsequent advancements, including extensions up to the sixth order by Konoplya \cite{Konoplya:2003ii} and up to the thirteenth order by Matyjasek and Opala \cite{Matyjasek:2017psv}, have further refined this approach.

In particular, this section is dedicated to exploring the \textit{quasinormal} modes specifically in the context of tensorial (gravitational) perturbations. Our focus will align closely with the findings of Ref. \cite{kim2004gravitational,araujo2024dark}. In general lines, the axially symmetric spacetime can be expressed as:
\begin{equation}
\mathrm{d}s^2 -e^{2\nu}\mathrm{d}t^2 + e^{2\psi}(\mathrm{d}\phi - q_1\mathrm{d}t - q_2\mathrm{d}r - q_3\mathrm{d}\theta )^2 e^{2\mu_2}\mathrm{d}r^2 + e^{2\mu_3}\mathrm{d}\theta^2. 
\end{equation}
Taking into account the non-perturbed black hole, we obtain 
\be e^{2\nu} = f(r), \quad
e^{-2\mu_2}=\left( 1 - \f{2m(r)}{r} \right) = \f{\Delta}{r^2}, 
\end{equation}
in which
\be
\Delta = r^2 - 2 m(r) r, \quad e^{\mu_3} = r, \quad e^\psi = r
\sin\theta, \ee and \be q_1=q_2=q_3=0, \ee
where the metric presented in \eqref{lalalaa} can properly be written as follows
\begin{eqnarray}
    f_{\Theta}(r)=1-\frac{2 \mathcal{M}_{\Theta}(r)}{r},
\end{eqnarray}
so that 
\begin{eqnarray}
\nonumber
    \mathcal{M}_{\Theta}(r) = M - \frac{4M \sqrt{\Theta}}{\sqrt{\pi}r}.
\end{eqnarray}

Axial perturbations are commonly characterized by $q_1$, $q_2$, and $q_3$. It is worth noting that for the linear perturbations $\delta\nu, \delta\psi, \delta\mu_2, \delta\mu_3$, polar ones with even parity emerge. However, such perturbations fall outside the scope of this paper. Now, turning our attention to Einstein's equations, we find
\be ( e^{3\psi+\nu-\mu_2-\mu_3}
Q_{23} )_{,3} = - e^{3\psi-\nu-\mu_2+\mu_3} Q_{02,0},\ee where
$x^2 = r, x^3 = \theta$ and $Q_{AB}=q_{A,B}-q_{B,A}, Q_{A0} =
q_{A,0}-q_{1,A}$ \cite{kim2004gravitational,araujo2024dark}. In addition, we can also rewrite the above expression as 
\be
\f{\sqrt{f(r)}}{\sqrt{\Delta}}\f{1}{r^3\sin^3\theta}\f{\pa
Q}{\pa\theta} = - (q_{1,2} - q_{2,0})_{,0},\ee in which $Q$ is given as \be
Q(t,r,\theta) = \Delta Q_{23}\sin^3\theta = \Delta
(q_{2,3}-q_{3,2})\sin^3\theta. \ee

Considering another significant equation, namely,
\be ( e^{3\psi+\nu-\mu_2-\mu_3} Q_{23} )_{,2} =
e^{3\psi-\nu+\mu_2-\mu_3} Q_{03,0}, \ee it can be shown that \be
\f{\sqrt{f(r)}\sqrt{\Delta}}{r^3\sin^3\theta}\f{\pa Q}{\pa\theta} =
(q_{1,3} - q_{3,0})_{,0}.\ee

We can demonstrate this further by employing the expression $Q(r,\theta) = Q(r)C^{-3/2}_{l+2}(\theta)$, where $C^\nu_n (\theta)$ denotes the Gegenbauer function that obeys
\cite{kim2004gravitational,araujo2024dark},
\be
\left[ \f{\mathrm{d}}{\mathrm{d}\theta}\sin^{2\nu}\theta \f{\mathrm{d}}{\mathrm{d}\theta} +
n(n+2\nu)\sin^{2\nu}\theta \right] C^\nu_n (\theta) = 0, \ee 
therefore
\be r \sqrt{f(r) \Delta} \f{\mathrm{d}}{\mathrm{d}r} \left( \f{\sqrt{f(r) \Delta}
}{r^3} \f{\mathrm{d}Q}{\mathrm{d}r} \right) - \mu^2
\f{f(r)}{r^2}Q + \omega^2Q = 0,  \ee 
where
$\mu^2=(l-1)(l+2)$. Here, we regard $Q=rZ$ so that $\f{\mathrm{d}}{\mathrm{d}r_*}=
\sqrt{f(r) \Delta}\f{1}{r}\f{\mathrm{d}}{\mathrm{d}r} $, and
\ie
\label{eqqq}
\left( \f{\mathrm{d}^2}{\mathrm{d}r^{*2}} +
\omega^2 - V_{eff}(r) \right) Z = 0, \fe
with $r^{*}$ being the tortoise coordinate. Explicitly, it reads
\ie
r^{*} = r + M \ln \left(8 \sqrt{\Theta } M+\sqrt{\pi } r (r-2 M)\right)+\frac{2 \sqrt{M} \left(\sqrt{\pi } M-4 \sqrt{\Theta }\right) \tanh ^{-1}\left(\frac{\sqrt[4]{\pi } (M-r)}{\sqrt{M} \sqrt{\sqrt{\pi } M-8 \sqrt{\Theta }}}\right)}{\sqrt[4]{\pi } \sqrt{\sqrt{\pi } M-8 \sqrt{\Theta }}}.
\fe
Furthermore, the effective potential can properly be obtained
\begin{equation}
    V_{eff}(r)=f_{\Theta}(r) \left(\frac{l(l+1)}{r^2} - \frac{6 \mathcal{M}_{\Theta}(r)}{r^3}+\frac{2 \mathcal{M}_{\Theta}'(r)}{r^2} \right),
\end{equation}
in such a way that 
\begin{equation}
    V_{eff}(r)=f_{\Theta}(r) \left(\frac{l(l+1)}{r^2}  +  \frac{2 M \left(\frac{16 \sqrt{\Theta }}{\sqrt{\pi }}-3 r\right)}{r^4} \right).
\end{equation}

\begin{figure}
    \centering
      \includegraphics[scale=0.42]{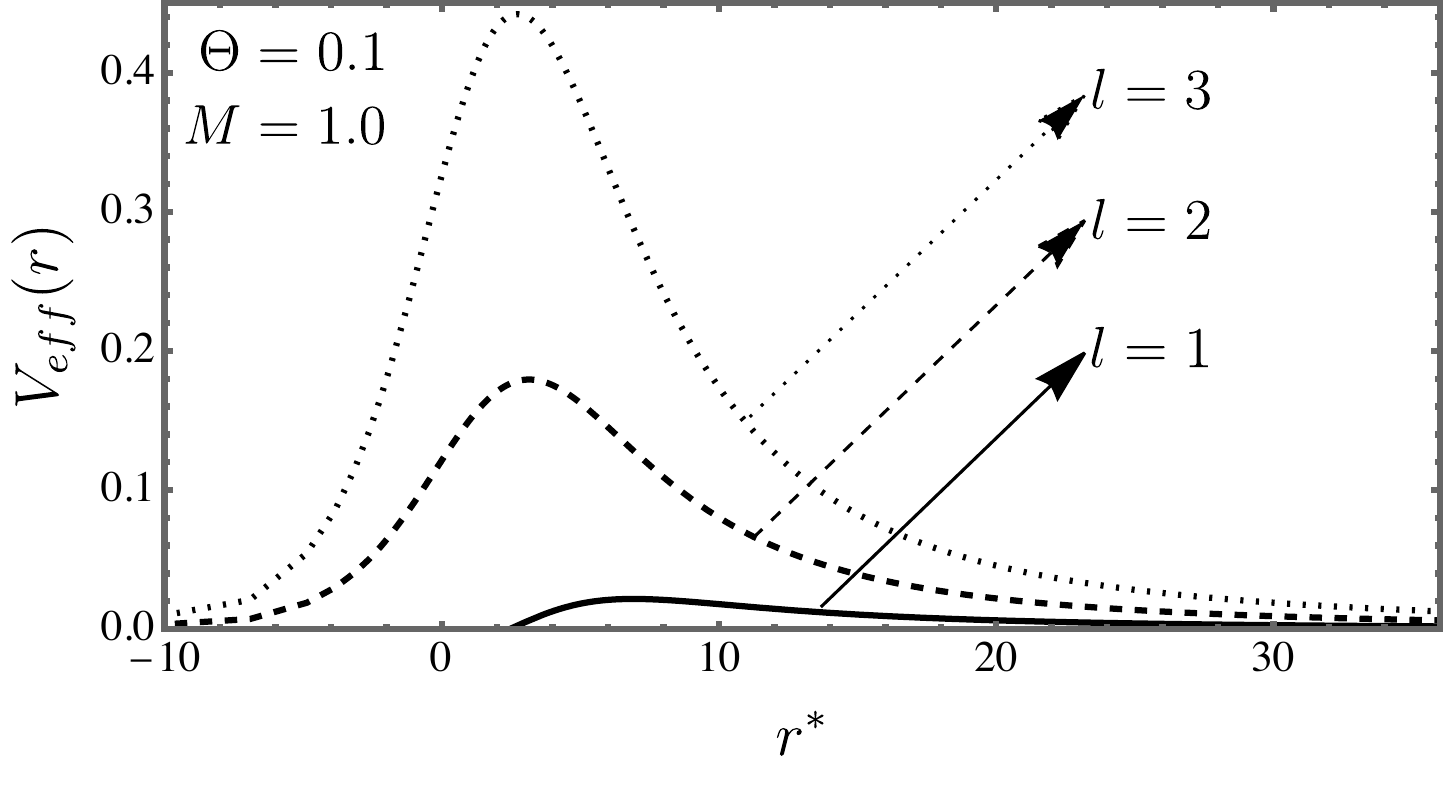}
    \caption{The effective potential $V_{eff}(r)$ is shown as a function of the tortoise coordinate $r^{*}$ for tensorial perturbations.}
    \label{effectivepotentialtensorial}
\end{figure}

As it is usually inferred in the studies of \textit{quasinormal} modes, we provide the plots of the effective potential $V_{eff}(r)$ against the tortoise coordinate $r^{*}$ in  Fig. \ref{effectivepotentialtensorial}. Here, the graphic is performed by considering particular values of $\Theta=0.1$ and $M=1.0$ for different values of $l$.

Effectively solving Eq. \eqref{eqqq} requires carefully considering the relevant boundary conditions. In our particular scenario, solutions stand out by demonstrating purely ongoing behavior near the horizon
\[
    Z^{\text{in}}(r^{*}) \sim 
\begin{cases}
    C_{l}(\omega) e^{-i\omega r^{*}} & ( r^{*}\rightarrow - \infty)\\
    A^{(-)}_{l}(\omega) e^{-i\omega r^{*}} + A^{(+)}_{l}(\omega) e^{+i\omega r^{*}} & (r^{*}\rightarrow + \infty).\label{boundaryconditions11}
\end{cases}
\]
The integral complex coefficients $A^{(+)}_{l}(\omega)$, $C_{l}(\omega)$, and $A^{(-)}_{l}(\omega)$ are considered crucial for our subsequent analytical investigation. They are utilized as fundamental elements in the examination of the \textit{quasinormal} modes associated with a black hole, distinguished by frequencies \(\omega_{nl}\), satisfying the condition $A^{(-)}_{l}(\omega_{nl})=0$. These modes are characterized by a distinctive behavior, with purely outgoing waves observed at spatial infinity and exclusively ongoing waves near the event horizon. The integers $n$ and $l$ are denoted as the overtone and multipole numbers, respectively. Furthermore, it is noteworthy that the spectrum of \textit{quasinormal} modes represents the eigenvalues derived from Eq. \eqref{eqqq}. In order to provide the calculation of these frequencies, the WKB method, a semi-analytical technique reminiscent of quantum mechanics, is invoked.

Moreover, the WKB approximation, which Schutz and Will introduced \cite{1985ApJ...291L..33S}, has become a remarkable method to address the \textit{quasinormal} modes, particularly in analyzing particle scattering around black holes. This technique has undergone refinement over the years, with significant contributions from Konoplya \cite{Konoplya:2003ii, Konoplya:2004ip}. However, it is crucial to recognize that the method's applicability relies on the potential exhibiting a barrier-like structure (as shown in Fig. \ref{effectivepotentialtensorial}) and leveling off to constant values as $r^{*} \to \pm \infty$. By aligning the solution power series with the peak potential turning points, researchers can accurately derive the \textit{quasinormal} modes \cite{Santos:2015gja}. Given these premises, the expression for the sixth-order WKB formula is as follows:
\begin{equation}
\frac{i(\omega^{2}_{n}-V_{0})}{\sqrt{-2 V^{''}_{0}}} - \sum^{6}_{j=2} \Lambda_{j} = n + \frac{1}{2}.
\end{equation}

In essence, Konoplya's formulation for the \textit{quasinormal} modes comprises several key components. In particular, the term $V^{''}_{0}$ represents the second derivative of the potential, evaluated at its maximum $r_{0}$. Furthermore, the constants $\Lambda_{j}$ are influenced by the effective potential and its derivatives at this peak. Notably, recent progress in this field has revealed a 13th-order WKB approximation developed by Matyjasek and Opala \cite{Matyjasek:2017psv}.

It is to be noted that the \textit{quasinormal} frequencies are characterized by a negative imaginary component. This feature implies that these modes are subject to exponential decay over time, indicating energy dissipation through gravitational waves. This observation is in line with previous studies examining scalar, electromagnetic, and gravitational perturbations within spherically symmetric contexts \cite{Konoplya:2011qq,Berti:2009kk,Heidari:2023bww,2023InJPh.tmp..228C,araujo2024dark}.

In Tab. \ref{table1}, we present the \textit{quasinormal} modes for a variety of values of $\Theta$; also, we have considered $l=0$, $l=1$, and $l=2$. In general lines, the non-commutative parameter is responsible for attenuating the dumped frequencies. Similar conclusions were addressed recently in the literature, concerning distinct methods to introduce the non-commutativity in gravity \cite{t25:1,t25:2}.

\begin{table*}[tbp]
\begin{tabular}{|l|l|l|l|l|}
\hline
 \multicolumn{1}{|c|}{ $\text{Spin 2}$ } &  \multicolumn{1}{c|}{  $l=0$ } & \multicolumn{1}{c|}{  $l=1$ } & \multicolumn{1}{c|}{  $l=2$ }\\\hline
   \,\,\, $\Theta$ & \hspace{1.1cm}$\omega_{0}$ & \hspace{1.1cm}$\omega_{0}$ & \hspace{1.1cm}$\omega_{0}$   \\ \hline
\, $0.10$ & 4.84572 - 5.88427$i$ & 4.18317 - 5.95134$i$ & 2.89163 - 6.04328$i$  \\ 
\, $0.12$ & 4.02939 - 4.82192$i$ & 3.42490 - 4.88470$i$ & 2.24718 - 4.97041$i$   \\
\, $0.14$ & 3.44481 - 4.07124$i$ & 2.88527 - 4.13141$i$ & 1.79907 - 4.20667$i$  \\ 
\, $0.16$ & 3.00574 - 3.51331$i$ & 2.48475 - 3.56838$i$ & 1.47109 - 3.63607$i$  \\ 
\, $0.18$ & 2.66628 - 3.08019$i$ & 2.17584 - 3.13252$i$  & 1.22090 - 3.19606$i$  \\ 
\, $0.20$ & 2.39581 - 2.73483$i$ & 1.93000 - 2.78633$i$ & 1.02532 - 2.84456$i$  
\\\hline
\end{tabular}
\caption{\label{table1} The \textit{quasinormal} modes are shown by using sixth-order WKB approximation, considering different values of $\Theta$. Here, the multipole numbers are regarded $l=0$, $l=1$, and $l=2$.}
\end{table*}

%%%%%%%%%%%%%%%%%%%%%%%%%%%%%%%%%%%%%%%%%%%%%%%%%%%%%%%%%%%%%%%%%%%%%%%%%%%%%%%%%%%%%%%%%%%%%%%%%%%%%%%%%%%%%%%%%%%%%%%%%%%%%%%%%%%%%%%%%%%%%%%%%%%%%%%%%%%%%%%%%%%%%%%%%%%%%%%%%%%%%%%%%%%%%%%%%%%%%%%%%%%%%%%%%%%%%%%%%%%%%%%%%%%%%%%%%%%%%%%%%%%%%%%%%%%%%%%%%%%%%%%%%%%%%%%%%%%%%%%%%%%%%%%%%%%%%%%%%%%%%%%%%%%%%%%%%%%%%%%%%%%%%%%%%%%%%%%%%%%%%%%%%%%%%%%%%%%%%%%%%%%%%%%%%%%%%%%%%%%%%%%%%%%%%%%%%%%%%%%%%%%%%%%%%%%%%%%%%%%%%%%%%%%%%%
\section{Weak deflection angle using Gauss--Bonnet theorem \label{weak}}

In this section, we review the Gauss-Bonnet theorem and compute the weak deflection angle of the black hole. Initially, we express the null geodesics satisfying $\mathrm{d}s^2=0$, a rearrangement of which yields:

\begin{eqnarray}
\mathrm{d}t^2=\gamma_{ij}\mathrm{d}x^i \mathrm{d}x^j=\frac{1}{f_{\Theta}(r)^2}\mathrm{d}r^2+\frac{r^2}{f_{\Theta}(r)}\mathrm{d}\Omega^2,~\label{opmetric}
\end{eqnarray}

Here, $i$ and $j$ range from $1$ to $3$, and $\gamma_{ij}$ represents the optical metric. 

To employ the Gauss-Bonnet theorem, it is imperative to compute the Gaussian curvature, and this calculation is performed here:
\begin{eqnarray}
\mathcal{K}=\frac{R}{2}&=& \frac{f_{\Theta}(r)}{2} \frac{\mathrm{d}^{2}}{\mathrm{d} r^{2}}f_{\Theta}(r) -\frac{\left(\frac{\mathrm{d}}{\mathrm{d} r}f_{\Theta}(r)\right)^{2}}{4}=\frac{128 \Theta  M^2}{\pi  r^6}-\frac{48 \sqrt{\Theta } M^2}{\sqrt{\pi } r^5}+\frac{3 M^2}{r^4}+\frac{24 \sqrt{\Theta } M}{\sqrt{\pi } r^4}-\frac{2 M}{r^3}.~\label{GC}
\end{eqnarray}
Here, $\gamma\equiv\det (\gamma_{ij})$, and $R$ represents the Ricci scalar. The surface area on the equatorial plane is expressed as \cite{Gibbons:2008rj}:

\begin{equation}
\mathrm{d}S=\sqrt\gamma \mathrm{d}r \mathrm{d} \phi= \frac{r}{f_{\Theta}(r)^{3/2}} \mathrm{d}r \mathrm{d}\phi \approx \left(r + \frac{3 M \left(\sqrt{\pi } r-4 \sqrt{\Theta }\right)}{\sqrt{\pi } r}\right) \mathrm{d}r \mathrm{d}\phi.
\label{dsss}
\end{equation}

%\begin{widetext}

Subsequently, the deflection angle of light can be computed as:

\begin{eqnarray}
\alpha&=&-\int\int_{\tilde{D}}\mathcal{K}\mathrm{d}S=-\int^{\pi}_0\int^{\infty}_{\frac{b}{\sin\phi}}\mathcal{K}\mathrm{d}S \nonumber\\
&\simeq&\frac{4 M}{b}+\frac{15 \Theta  M^2}{b^4}-\frac{64 \sqrt{\Theta } M^2}{3 \sqrt{\pi } b^3}+\frac{3 \pi  M^2}{4 b^2}-\frac{6 \sqrt{\pi } \sqrt{\Theta } M}{b^2}.~\label{deflang}
\end{eqnarray}

\begin{figure}
    \centering
\includegraphics[width=0.5\linewidth]{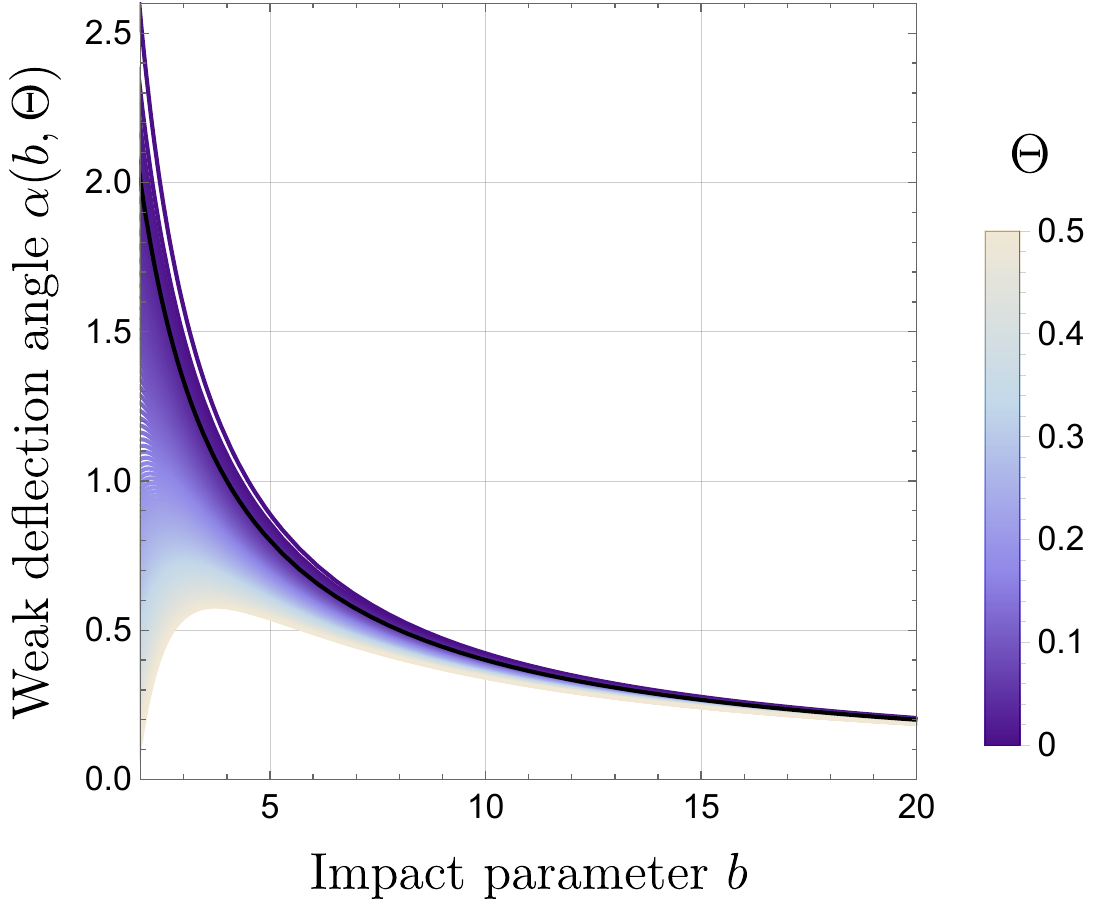}
    \caption{Weak deflection angle $\alpha$ versus impact parameter $b$ for $M=1$ and variable $\Theta$. The black solid line is for the Schwarzschild case.}
    \label{FigLensing}
\end{figure}

In this calculation, the zero-order particle trajectory $r=b/\sin\phi$, where $0\leq\phi\leq\pi$ in the weak deflection limit, has been employed which is shown in Fig. \ref{FigLensing}. As 
$\Theta$ increases from 0 to 0.5, the deflection angle 
$\alpha$ decreases. This suggests that higher values of $\Theta$ reduce the deflection angle, indicating a weaker gravitational lensing effect. For larger $b$ values, the deflection angle is predominantly influenced by the leading term $\frac{4 M}{b}$. The differences due to varying $\Theta$ become less pronounced in this range. At smaller $b$ values, the higher-order terms involving $\Theta$ and $M^2$ become more significant, resulting in noticeable differences in the deflection angle for different $\Theta$ values. The plot effectively illustrates the impact of 
$\Theta$ on the deflection angle, showing that higher 
$\Theta$ values lead to a smaller deflection angle for light passing near the black hole.

%%%%%%%%%%%%%%%%%%%%%%%%%%%%%%%%%%%%%%%%%%%%%%%%%%%%%%%%%%%%%%%%%%%%%%%%%%%%%%%%%%%%%%%%%%%%%%%%%%%%%%%%%%%%%%%%%%%%%%%%%%%%%%%%%%%%%%%%%%%%%%%%%%%%%%%%%%%%%%%%%%%%%%%%%%%%%%%%%%%%%%%%%%%%%%%%%%%%%%%%%%%%%%%%%%%%%%%%%%%%%%%%%%%%%%%%%%%%%%%%%%%%%%%%%%%%%%%%%%%%%%%%%%%%%%%%%%%%%%%%%%%%%%%%%%%%%%%%%%%%%%%%%%%%%%%%%%%%%%%%%%%%%%%%%%%%%%%%%%%%%%%%%%%%%%%%%%%%%%%%%%%%%%%%%%%%%%%%%%%%%%%%%%%%%%%%%%%%%%%%%%%%%%%%%%%%%%%%%%%%

\section{Gravitational lensing in the strong deflection limit \label{strong}}

In this section, we describe the general methodology used to obtain the deflection angle of a light ray within the strong deflection limit \cite{tsukamoto2017deflection}. Like many previous works (e.g., \cite{Nascimento:2020ime,nascimento2024gravitational}), we focus on asymptotically flat, static, and spherically symmetric spacetimes characterized by the line element:
\ie
\label{ssst}
\mathrm{d}s^{2} = - A(r) \mathrm{d}t^{2} + B(r) \mathrm{d}r^{2} + C(r)(\mathrm{d}\theta^2 + \sin^{2}\theta\mathrm{d}\phi^2).
\fe
To apply the method proposed by Tsukamoto \cite{tsukamoto2017deflection}, the metric must satisfy the asymptotic flatness condition. Specifically, the coefficients $A(r)$, $B(r)$, and $C(r)$ should behave as follows: $\lim\limits_{r \to \infty}A(r) = 1$, $\lim\limits_{r \to \infty}B(r) = 1$, and $\lim\limits_{r \to \infty}C(r) = r^{2}$. Due to the spacetime symmetries, there are two Killing vectors: $\partial_t$ and $\partial_\phi$. 

We shall now briefly describe the procedure for calculating the deflection angle in the strong field regime, starting by defining a new variable, $D(r)$:
\ie
D(r) \equiv \frac{C^{\prime}(r)}{C(r)} - \frac{A^{\prime}(r)}{A(r)},
\fe
with the prime indices representing the derivative with respect to the radial coordinate. We assume there is at least one positive solution when \(D(r) = 0\). The radius of the photon sphere, denoted as \(r_m\), is defined as the largest positive solution of \(D(r) = 0\). We assume \(A(r)\), \(B(r)\), and \(C(r)\) are finite and positive for \(r \geq r_m\).

Because of the existence of two Killing vectors, there are two conserved quantities: energy \(E = A(r)\Dot{t}\) and angular momentum \(L = C(r)\Dot{\phi}\). We assume both \(E\) and \(L\) are nonzero. With this, we introduce the impact parameter \(b\), defined as:
\ie
b \equiv \frac{L}{E} = \frac{C(r)\Dot{\phi}}{A(r)\Dot{t}}.
\fe
Due to axial symmetry, we can restrict the motion to the equatorial plane (\(\theta = \pi/2\)), so the radial equation becomes:
\ie
\Dot{r}^{2} = V(r),
\fe
with \(V(r) = L^{2} R(r)/B(r)C(r)\), and \(R(r) \equiv C(r)/A(r) b^{2} - 1\). This equation is identical to the motion equation for a unit mass object in a potential \(V(r)\). The photon can move where \(V(r) \geq 0\). Given the asymptotic flatness conditions, \(\lim\limits_{{r \to \infty}} V(r) = E^{2} > 0\), allowing the photon to exist at infinity (\(r \to \infty\)). We assume there is at least one positive solution for \(R(r) = 0\).

In this study, we focus on the gravitational lensing scenario where a photon, originating from infinity, approaches a gravitational object, scatters at the closest distance \(r_0\), and then continues to infinity. For scattering to occur, \(r_m < r_0\) must hold, as a photon cannot have a closed orbit in this case. Here, \(r = r_0\) is the largest positive solution of \(R(r) = 0\), and \(B(r)\) and \(C(r)\) are finite. Consequently, \(V(r)\) vanishes at \(r = r_0\). Since \(r_0\) is the point of closest approach where \(R(r) = 0\), we deduce:
\ie
A_{0}\Dot{t}^{2}_{0} = C_{0}\Dot{\phi}^{2}_{0}.
\fe

In this manner, and throughout the following discussions, the subscript \( ``0" \) signifies quantities evaluated at \(r = r_0\). For simplicity, we can assume, without loss of generality, that the impact parameter \(b\) is positive, especially when dealing with a single light ray. As the impact parameter remains constant along the trajectory, it can be expressed as:
\ie
b(r_{0}) = \frac{L}{E} = \frac{C_{0}\Dot{\phi}_{0}}{A_{0}\Dot{t}_{0}} = \sqrt{\frac{C_{0}}{A_{0}}}.
\fe
Notice that $R(r)$ may also be given by
\ie
R(r)= \frac{A_{0}C}{AC_{0}} - 1.
\fe

We outline a condition that is both necessary and sufficient for the existence of a circular light orbit, drawing from the approach as shown in Ref. \cite{hasse2002gravitational}. Therefore, the trajectory equation reads
\ie
\frac{BC \Dot{r}^{2}}{E^{2}} + b^{2} = \frac{C}{A},
\fe
in such a way that
\ie
\ddot{r} + \frac{1}{2}\left( \frac{B^{\prime}}{B} + \frac{C^{\prime}}{C} \Dot{r}^{2} \right) = \frac{E^{2}D(r)}{AB}. 
\fe
Here, for \(r \geq r_m\), \(A(r)\), \(B(r)\), and \(C(r)\) are finite and positive. With \(E\) being positive as well, the condition \(D(r)=0\) ensures the consistency of a circular light orbit. It is worth noting that \(R^{\prime}_{m} = D_{m}C_{m}A_{m}/b^{2} = 0\), where the subscript \(m\) denotes quantities specifically evaluated at \(r = r_m\).

Now, we consider a critical impact parameter, denoted by \(b_c\):
\ie
b_{c}(r_{m}) \equiv \lim_{r_{0} \to r_{m}} \sqrt{\frac{C_{0}}{A_{0}}}.
\fe
This regard will henceforth be referred to as the strong deflection limit. By taking a derivative of \(V(r)\) with respect to \(r\), we have
\ie
V^{\prime}(r) = \frac{L^{2}}{BC} \left[ R^{\prime} + \left( \frac{C^{\prime}}{C} - \frac{B^{\prime}}{B}   \right)   R  \right].
\fe
This means that as \(r_{0}\) approaches \(r_{m}\) in the strong deflection limit, both \(V(r_{0})\) and \(V^{\prime}(r_{0})\) tend to zero. Consequently, the trajectory equation adopts the following form:
\ie
\left(  \frac{\mathrm{d}r}{\mathrm{d}\phi}     \right)^{2} = \frac{R(r)C(r)}{B(r)},
\fe
and the so-called the deflection  angle, i.e., $\alpha(r_{0})$, can be given by
\ie
\alpha(r_{0}) = I(r_{0}) - \pi,
\fe
where, in this context, $I(r_{0})$ is given by
\ie
I(r_{0}) \equiv 2 \int^{\infty}_{r_{0}} \frac{\mathrm{d}r}{\sqrt{\frac{R(r)C(r)}{B(r)}}}.
\fe

To proceed, our initial task involves taking into account the integration. It is important to mention that such a procedure poses a formidable challenge, as noted in the work by Tsukamoto \cite{tsukamoto2017deflection}. In addition, let us define \cite{tsukamoto2017deflection}
\ie
z \equiv 1 - \frac{r_{0}}{r},
\fe
so that the integral is rewritten as
\ie
I(r_{0}) = \int^{1}_{0} f(z,r_{0}) \mathrm{d}z,
\fe
with 
\ie
f(z,z_{0}) \equiv \frac{2r_{0}}{\sqrt{G(z,r_{0})}}, \,\,\,\,\,\,\,\, \text{and} \,\,\,\,\,\,\,\,  G(z,r_{0}) \equiv R \frac{C}{B}(1-z)^{4}.
\fe

In terms of $z$, notice that $R(r)$ reads
\ie
R(r) = D_{0}r_{0} z + \left[ \frac{r_{0}}{2}\left( \frac{C^{\prime\prime}_{0}}{C_{0}} - \frac{A_{0}^{\prime\prime}}{A_{0}}  \right) + \left( 1 - \frac{A_{0}^{\prime}r_{0}}{A_{0}}  \right) D_{0}  \right] r_{0} z^{2} + \mathcal{O}(z^{3})+ ...    \,\,\,\,.
\fe
Expanding $G(z,r_{0})$ in terms of $z$, we have:
\ie
G(z,r_{0}) = \sum^{\infty}_{n=1} c_{n}(r_{0})z^{n},
\fe
where $c_{1}(r)$ and $c_{2}(r)$ are
\ie
c_{1}(r_{0}) = \frac{C_{0}D_{0}r_{0}}{B_{0}},
\fe
and
\ie
c_{2}(r_{0}) = \frac{C_{0}r_{0}}{B_{0}} \left\{ D_{0} \left[ \left( D_{0} - \frac{B^{\prime}_{0}}{B_{0}}  \right)r_{0} -3       \right] + \frac{r_{0}}{2} \left(  \frac{C^{\prime\prime}_{0}}{C_{0}} - \frac{A^{\prime\prime}_{0}}{A_{0}}  \right)                 \right\}.
\fe

Moreover, by considering the strong deflection limit, it turns out that
\ie
c_{1}(r_{m}) = 0, \,\,\,\,\,\, \text{and} \,\,\,\,\,\, c_{2}(r_{m}) =  \frac{C_{m}r^{2}_{m}}{2 B_{m}}D^{\prime}_{m}, \,\,\,\,\,\,\, \text{with} \,\,\,\,\, D^{\prime}_{m} = \frac{C^{\prime\prime}}{C_{m}} - \frac{A^{\prime\prime}}{A_{m}}.
\fe
Here, $G(z,r_{0})$ possess a shorter notation, as shown below:
\ie
G_{m}(z) = c_{2}(r_{m})z^{2} + \mathcal{O}(z^{3}).
\fe

This shows that the main divergence of \(f(z,r_0)\) happens at the order of \(z^{-1}\), causing a logarithmic divergence in the integral \(I(r_0)\) as \(r_0\) gets close to \(r_m\). To manage this divergence, we split the integral \(I(r_0)\) into two parts: a divergent part, \(I_D(r_0)\), and a regular part, \(I_R(r_0)\). In this sense, the divergent part \(I_D(r_0)\) is properly written as:
\ie
I_{D}(r_{0}) \equiv \int^{1}_{0} f_{D}(z,r_{0}) \mathrm{d}z, \,\,\,\,\,\,\, \text{with} \,\,\,\,\,\,f_{D}(z,r_{0}) \equiv \frac{2 r_{0}}{\sqrt{c_{1}(r_{0})z + c_{2}(r_{0})z^{2}}}.
\fe
After integrating, we obtain
\ie
I_{D} (r_{0}) = \frac{4 r_{0}}{\sqrt{c_{2}(r_{0})}} \ln \left[  \frac{\sqrt{c_{2}(r_{0})} + \sqrt{c_{1}(r_{0}) + c_{2}(r_{0})     }  }{\sqrt{c_{1}(r_{0})}}  \right].
\fe

Considering the expansion of \(c_{1}(r_{0})\) and \(b(r_{0})\) around \(r_{0} - r_{m}\)
\ie
c_{1}(r_{0}) = \frac{C_{m}r_{m}D^{\prime}_{m}}{B_{m}} (r_{0}-r_{m}) + \mathcal{O}((r_{0}-r_{m})^{2}),
\fe
and
\ie
b(r_{0}) = b_{c}(r_{m}) + \frac{1}{4} \sqrt{\frac{C_{m}}{A_{m}}}D^{\prime}_{m}(r_{0}-r_{m})^{2} + \mathcal{O}((r_{0}-r_{m})^{3}),
\fe
leading to the following in the strong deflection limit
\ie
\lim_{r_{0} \to r_{m}} c_{1}(r_{0})  =  \lim_{b \to b_{c}} \frac{2 C_{m} r_{m} \sqrt{D^{\prime}}}{B_{m}} \left(  \frac{b}{b_{c}} -1  \right)^{1/2}.
\fe
Thereby, $I_{D}(b)$ is
\ie
I_{D}(b) = - \frac{r_{m}}{\sqrt{c_{2}(r_{m})}} \ln\left[ \frac{b}{b_{c}} - 1 \right] + \frac{r_{m}}{\sqrt{c_{2}(r_{m})}}\ln \left[ r^{2}D^{\prime}_{m}\right] + \mathcal{O}[(b-b_{c})\ln(b-b_{c})].
\fe

Meanwhile, we define $I_{R}(b)$ as
\ie
I_{R}(b) = \int^{0}_{1} f_{R}(z,b_{c})\mathrm{d}z + \mathcal{O}[(b-b_{c})\ln(b-b_{c})].
\fe
Here, let \( f_{R} \) be defined as \( f_{R} = f(z, r_{0}) - f_{D}(z, r_{0}) \). In the strong deflection limit, the deflection angle is given by
\ie
a(b) = - \Tilde{a} \ln \left[ \frac{b}{b_{c}}-1    \right] + \Tilde{b} + \mathcal{O}[(b-b_{c})\ln(b-b_{c})],
\label{deflections}
\fe
in which
\ie
\Tilde{a} = \sqrt{\frac{2 B_{m}A_{m}}{C^{\prime\prime}_{m}A_{m} - C_{m}A^{\prime\prime}_{m}}}, \,\,\,\,\,\,\,\, \text{and} \,\,\,\,\,\,\,\, \Tilde{b} = \Tilde{a} \ln\left[ r^{2}_{m}\left( \frac{C^{\prime\prime}}{C_{m}}  -  \frac{A^{\prime\prime}_{m}}{C_{m}} \right)   \right] + I_{R}(r_{m}) - \pi.
\fe

%%%%%%%%%%%%%%%%%%%%%%%%%%%%%%%%%%%%%%%%%%%%%%%%%%%%%%%%%%%%%%%%%%%%%%%%%%%%%%%%%%%%%%%%%%%%%%%%%%%%%%%%%%%%%%%%%%%%%%%%%%%%%%%%%%%%%%%%%%%%%%%%%%%%%%%%%%%%%%%%%%%%%%%%%%%%%%%%%%%%%%%%%%%%%%%%%%%%%%%%%%%%%%%%%%%%%%%%%%%%%%%%%%%%%%%%%%%%%%%%%%%%%%%%%%%%%%%%%%%%%%%%%%%%%%%%%%%%%%%%%%%%%%%%%%%%%%%%%%%%%%%%%%%%%%%%%%%%%%%%%%%%%%%%%%%%%%%%%%%%%%%%%%%%%%%%%%%%%%%%%%%%%%%%%%%%%%%%%%%%%%%%%%%

\subsection{Gravitational lensing of a non-commutative black hole}

After all the methodology presented above, let us apply it to our metric under consideration in Eq. (\ref{lalalaa}). Then, we have
\ie
b_{c} = \frac{1}{2} \sqrt{\frac{\left(\sqrt{M \left(9 \sqrt{\pi } M-64 \sqrt{\Theta }\right)}+3 \sqrt[4]{\pi } M\right)^4}{6 \pi  M^2-32 \sqrt{\pi } \sqrt{\Theta } M+2 \pi ^{3/4} M \sqrt{M \left(9 \sqrt{\pi } M-64 \sqrt{\Theta }\right)}}} .
\fe
Also, $\Tilde{a}$ and $\Tilde{b}$ can be expressed as
\ie
\Tilde{a} = \sqrt[4]{\pi } \sqrt{\frac{1}{\sqrt{\pi }-\frac{64 \sqrt{\Theta } M}{\left(\frac{\sqrt{M \left(9 \sqrt{\pi } M-64 \sqrt{\Theta }\right)}}{\sqrt[4]{\pi }}+3 M\right)^2}}},
\fe
In this sense, we can write
\ie
\begin{split}
\Tilde{b} = \sqrt[4]{\pi } \sqrt{\frac{1}{\sqrt{\pi }-\frac{64 \sqrt{\Theta } M}{\left(\frac{\sqrt{M \left(9 \sqrt{\pi } M-64 \sqrt{\Theta }\right)}}{\sqrt[4]{\pi }}+3 M\right)^2}}} \ln \left(\frac{8}{\frac{\sqrt[4]{\pi } M}{\sqrt{M \left(9 \sqrt{\pi } M-64 \sqrt{\Theta }\right)}}+1}\right)
+ I_{R}(r_{m}) - \pi.
\end{split}
\fe

In contrast with what happens in the Schwarzschild case, notice that the contribution to the parameter $\Tilde{a}$ is fundamentally due to the feature coming from the non-commutativity. In addition, $I_{R}(r_{m})$ can be calculated as 
\ie
\begin{split}
 & I_{R}(r_{m}) =   \int_{0}^{1} \mathrm{d}z \left\{   \sqrt{2} \left(\sqrt{M \left(9 \sqrt{\pi } M-64 \sqrt{\Theta }\right)}+3 \sqrt[4]{\pi } M\right)  \right. \\
 & \left. \times  \left[  -\frac{1}{\sqrt{M z^2 \left(-64 \sqrt{\Theta }+3 \sqrt[4]{\pi } \sqrt{M \left(9 \sqrt{\pi } M-64 \sqrt{\Theta }\right)}+9 \sqrt{\pi } M\right)}}  \right.\right. \\
 & \left. \left.   +  \frac{1}{\sqrt{M z^2 \left(\sqrt[4]{\pi } (3-2 z) \sqrt{M \left(9 \sqrt{\pi } M-64 \sqrt{\Theta }\right)}+3 \sqrt{\pi } M (3-2 z)-16 \sqrt{\Theta } (z-2)^2\right)}}   \right] \right\} \\
 & = 0.82439.
\end{split}
\fe

Here, we have evaluated $I_{R}(r_{m})$ numerically, considering $M=2$ and $\Theta=0.1$. In this manner, the deflection angle shown in Eq. (\ref{deflections})  reads
\ie
\begin{split}
a(b) = &  -  \sqrt[4]{\pi } \sqrt{\frac{1}{\sqrt{\pi }-\frac{64 \sqrt{\Theta } M}{\left(\frac{\sqrt{M \left(9 \sqrt{\pi } M-64 \sqrt{\Theta }\right)}}{\sqrt[4]{\pi }}+3 M\right)^2}}}  \ln\left[  \frac{b}{\frac{1}{2} \sqrt{\frac{\left(\sqrt{M \left(9 \sqrt{\pi } M-64 \sqrt{\Theta }\right)}+3 \sqrt[4]{\pi } M\right)^4}{6 \pi  M^2-32 \sqrt{\pi } \sqrt{\Theta } M+2 \pi ^{3/4} M \sqrt{M \left(9 \sqrt{\pi } M-64 \sqrt{\Theta }\right)}}}}  - 1  \right]  \\
& + \sqrt[4]{\pi } \sqrt{\frac{1}{\sqrt{\pi }-\frac{64 \sqrt{\Theta } M}{\left(\frac{\sqrt{M \left(9 \sqrt{\pi } M-64 \sqrt{\Theta }\right)}}{\sqrt[4]{\pi }}+3 M\right)^2}}} \ln \left(\frac{8}{\frac{\sqrt[4]{\pi } M}{\sqrt{M \left(9 \sqrt{\pi } M-64 \sqrt{\Theta }\right)}}+1}\right) + 0.82439  - \pi \\
& + \mathcal{O}\left\{ \left(b - \frac{1}{2} \sqrt{\frac{\left(\sqrt{M \left(9 \sqrt{\pi } M-64 \sqrt{\Theta }\right)}+3 \sqrt[4]{\pi } M\right)^4}{6 \pi  M^2-32 \sqrt{\pi } \sqrt{\Theta } M+2 \pi ^{3/4} M \sqrt{M \left(9 \sqrt{\pi } M-64 \sqrt{\Theta }\right)}}} \right) \right.\\ 
&  \left. \times  \ln \left[ b - \frac{1}{2} \sqrt{\frac{\left(\sqrt{M \left(9 \sqrt{\pi } M-64 \sqrt{\Theta }\right)}+3 \sqrt[4]{\pi } M\right)^4}{6 \pi  M^2-32 \sqrt{\pi } \sqrt{\Theta } M+2 \pi ^{3/4} M \sqrt{M \left(9 \sqrt{\pi } M-64 \sqrt{\Theta }\right)}}}\right]\right\}.
\end{split}
\fe

\begin{figure}
    \centering
     \includegraphics[scale=0.4]{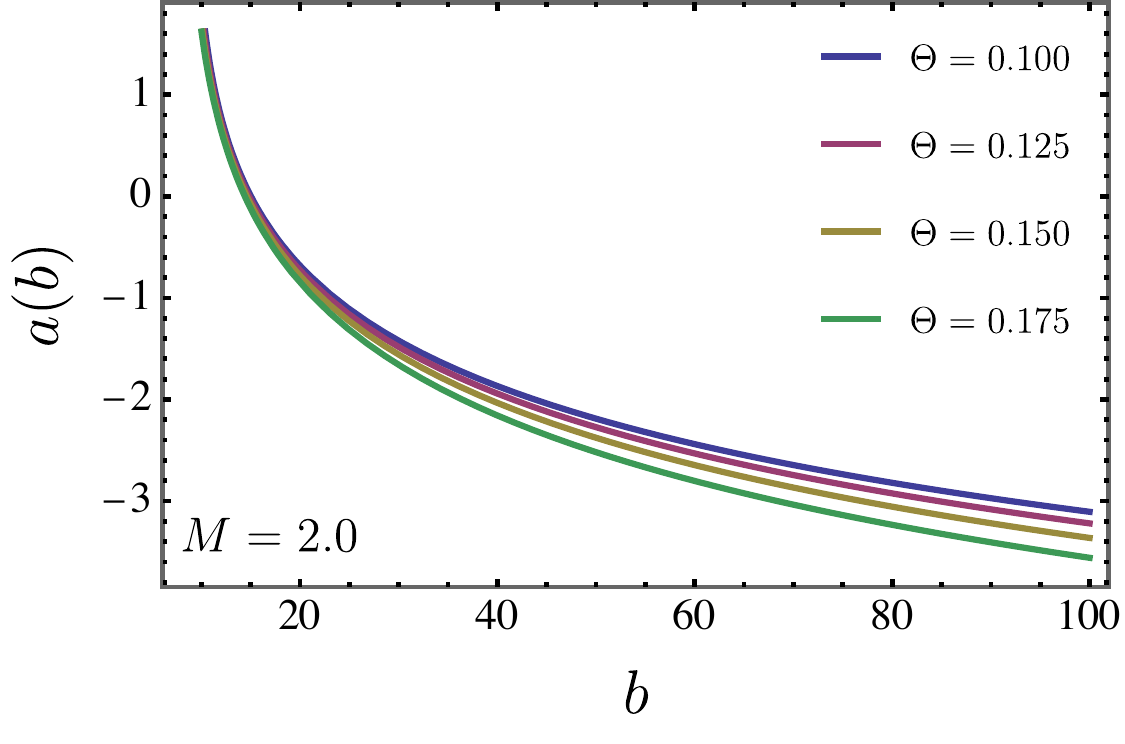}
    \includegraphics[scale=0.4]{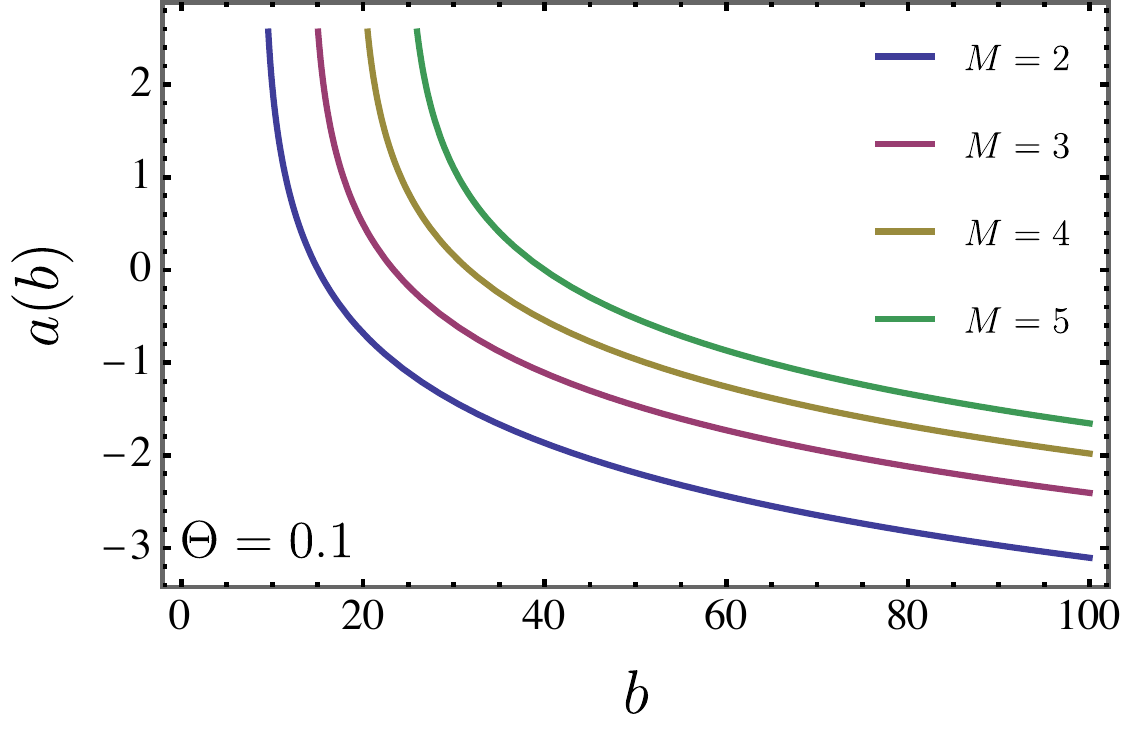}
    \caption{The deflection angle as a function of $b$ for different values of $\Theta$ and $M$.}
    \label{plotsdeflection}
\end{figure}

To enhance clarity for the reader, we present Fig. \ref{plotsdeflection}, illustrating the deflection angle as a function of $b$ across various system configurations.

\section{Conclusion\label{conclu}}

Our investigation was centered on exploring the gravitational signatures of a non-commutative black hole. Specifically, we selected a distinct mass distribution, described by the mathematical expression $\rho_{\Theta}(r) = M \sqrt{\Theta}/ \pi^{3/2}(r^{2}+\pi \Theta)^{2}$, to derive the corresponding black hole solution. This theory revealed the presence of two horizons: an event horizon and a Cauchy horizon.

In addition, we calculated the \textit{Hawking} temperature using two different methodologies: the topological method and surface gravity analysis. Additionally, we examined entropy and heat capacity to investigate further the thermodynamic properties of the system. Our comparisons extended to contrasting thermal quantities with those of the usual Schwarzschild black hole and its modified counterpart within non-commutative gauge theory \cite{t29}.

In our study of the evaporation process, we initially determined the remnant mass by considering the black hole's final state, where $T_{\Theta} \to 0$. Subsequently, we derived an analytical expression for the time evaporation up to a \textit{grey-body} factor. Furthermore, we estimated the black hole lifetime for specific initial and final masses, yielding $t_{evap} = \frac{1}{\xi} \left( 2.51807\times 10^7 \right)$. Furthermore, we computed the emission rate, taking into account the high-energy absorption cross-section for observers at infinity. Also, to address quasinormal modes, we analyzed tensorial perturbations using the 6th WKB approximation. The weak and strong deflection limits were computed by employing the Gauss-Bonnet theorem and Tsukamoto methods, respectively.

In the future, we aim to explore the gravitational lensing of black holes within a non-commutative gauge theory, which will expand our comprehension of these celestial phenomena.

%%%%%%%%%%%%%%%%%%%%%%%%%%%%%%%%%%%%%%%%%%%%%%%%%%%%%%%%%%%%%%%%%%%%%%%%%%%%%%%%%%%%%%%%%%%%%%%%%%%%%%%%%%%%%%%%%%%%%%%%%%%%%%%%%%%%%%%%%%%%%%%%%%%%%%%%%%%%%%%%%%%%%%%%%%%%%%%%%%%%%%%%%%%%%%%%%%%%%%%%%%%%%%%%%%%%%%%%%%%%%%%%%%%%%%%%%%%%%%%%%%%%%%%%%%%%%%%%%%%%%%%%%%%%%%%%%%%%%%%%%%%%%%%%%%%%%%%%%%%%%%%%%%%%%%%%%%%%%%%%%%%%%%%%%%%%%%%%%%%%%%%%%%%%%%%%%%%%%%%%%%%%%%%%%%%%%%%%%%%%%%%%%%%%%%%%%%%%%%%%%%%%%%%%%%%%%%%%%%%%%%%%%%%%%%%%%%%%%%%%%%%%%%%%%%%%%%%%%%%%%%%%%%%%%%%%%%%%%%%%%%%%%%%%%%%%%%%%%%%%%%%%

%%%%%%%%%%%%%%%%%%%%%%%%%%%%%%%%%%%%%%%%%%%%%%%%%%%%%%%%%%%%%%%%%%%%%%%%%%%%%%%%%%%%%%%%%%%%%%%%%%%%%%%%%%%%%%%%%%%%%%%%%%%%%%%%%%%%%%%%%%%%%%%%%%%%%%%%%%%%%%%%%%%%%%%%%%%%%%%%%%%%%%%%%%%%%%%%%%%%%%%%%%%%%%%%%%%%%%%%%%%%%%%%%%%%%%%%%%%%%%%%%%%%%%%%%%%%%%%%%%%%%%%%%%%%%%%%%%%%%%%%%%%%%%%%%%%%%%%%%%%%%%%%%%%%%%%%%%%%%%%%%%%%%%%%%%%%%%%%%%%%%%%%%%%%%%%%%%%%%%%%%%%%%%%%%%%%%%%%%%%%%%%%%%%%%%%%%%%%%%%%%%%%%%%%%%%%%%%%%%%%%%%%%%%%%%%%%%%%%%%%%%%%%%%%%%%%%%%%%%%%%%%%%%%%%%%%%%%%%%%%%%%%%%%%%%%%%%%%%%%%%%%%

%%%%%%%%%%%%%%%%%%%%%%%%%%%%%%%%%%%%%%%%%%%%%%%%%%%%%%%%%%%%%%%%%%%%%%%%%%%%%%%%%%%%%%%%%%%%%%%%%%%%%%%%%%%%%%%%%%%%%%%%%%%%%%%%%%%%%%%%%%%%%%%%%%%%%%%%%%%%%%%%%%%%%%%%%%%%%%%%%%%%%%%%%%%%%%%%%%%%%%%%%%%%%%%%%%%%%%%%%%%%%%%%%%%%%%%%%%%%%%%%%%%%%%%%%%%%%%%%%%%%%%%%%%%%%%%%%%%%%%%%%%%%%%%%%%%%%%%%%%%%%%%%%%%%%%%%%%%%%%%%%%%%%%%%%%%%%%%%%%%%%%%%%%%%%%%%%%%%%%%%%%%%%%%%%%%%%%%%%%%%%%%%%%%%%%%%%%%%%%%%%%%%%%%%%%%%%%%%%%%%
\pagebreak

\section*{Appendix\label{appendix}}

Here, we present the analytical expression for the evaporation time $t_{\text{evap}}$ up to a \textit{grey-body} factor: 

\ie
\begin{split}
& t_{\text{evap}} = \frac{8 \pi ^{7/2}}{\xi} \left\{  \frac{8}{243 \pi \delta M \left(\sqrt{\pi } \delta M-8 \sqrt{\Theta }\right)}  \left[  243 \pi ^2 \delta M^5+243 \pi ^{7/4} \delta M^4 \sqrt{\delta M \left(\sqrt{\pi } \delta M-8 \sqrt{\Theta }\right)}  \right.\right. \\
& \left.\left. + 81 \pi ^{7/4} \delta M^4 \sqrt{\delta M \left(9 \sqrt{\pi } \delta M-64 \sqrt{\Theta }\right)}                    +81 \pi ^{3/2} \delta M^3 \sqrt{\delta M \left(9 \sqrt{\pi } \delta M-64 \sqrt{\Theta }\right)} \sqrt{\delta M \left(\sqrt{\pi } \delta M-8 \sqrt{\Theta }\right)}   \right.\right.\\
& \left.\left. +180 \pi ^{5/4} \delta M^3 \sqrt{\Theta } \sqrt{\delta M \left(9 \sqrt{\pi } \delta M-64 \sqrt{\Theta }\right)}-972 \pi ^{3/2} \delta M^4 \sqrt{\Theta } \right.\right.\\
& \left.\left.
+504 \pi  \delta M^2 \sqrt{\Theta } \sqrt{\delta M \left(9 \sqrt{\pi } \delta M-64 \sqrt{\Theta }\right)} \sqrt{\delta M \left(\sqrt{\pi } \delta M-8 \sqrt{\Theta }\right)}    +13608 \pi  \delta M^3 \Theta       \right.\right. \\
&  \left.\left.  15552 \pi ^{3/4} \delta M^2 \Theta  \sqrt{\delta M \left(\sqrt{\pi } \delta M-8 \sqrt{\Theta }\right)}+5880 \pi ^{3/4} \delta M^2 \Theta  \sqrt{\delta M \left(9 \sqrt{\pi } \delta M-64 \sqrt{\Theta }\right)}   \right.\right.  \\
& \left. \left.    +8544 \sqrt{\pi } \delta M \Theta  \sqrt{\delta M \left(9 \sqrt{\pi } \delta M-64 \sqrt{\Theta }\right)} \sqrt{\delta M \left(\sqrt{\pi } \delta M-8 \sqrt{\Theta }\right)}-171072 \sqrt{\pi } \delta M^2 \Theta ^{3/2}    \right.\right. \\
&  \left.\left.       -373248 \sqrt[4]{\pi } \delta M \Theta ^{3/2} \sqrt{\delta M \left(\sqrt{\pi } \delta M-8 \sqrt{\Theta }\right)}-115584 \sqrt[4]{\pi } \delta M \Theta ^{3/2} \sqrt{\delta M \left(9 \sqrt{\pi } \delta M-64 \sqrt{\Theta }\right)}     \right.\right. \\
& \left. \left.  -266496 \Theta ^{3/2} \sqrt{\delta M \left(9 \sqrt{\pi } \delta M-64 \sqrt{\Theta }\right)} \sqrt{\delta M \left(\sqrt{\pi } \delta M-8 \sqrt{\Theta }\right)}  -124416 \delta M \Theta ^2     \right.\right.   \\
& \left.\left.   1813504 \delta M \Theta ^2 \ln (\delta M)-226688 \sqrt{\pi } \delta M^2 \Theta ^{3/2} \ln (\delta M)  \right.\right.\\ 
& \left.\left.   -217728 \sqrt{\pi } \delta M^2 \Theta ^{3/2} \ln \left(\sqrt[4]{\pi } \sqrt{\delta M \left(9 \sqrt{\pi } \delta M-64 \sqrt{\Theta }\right)}+5 \sqrt{\pi } \delta M-32 \sqrt{\Theta }\right)     \right.\right.\\
& \left.\left.   +1741824 \delta M \Theta ^2 \ln \left(\sqrt[4]{\pi } \sqrt{\delta M \left(9 \sqrt{\pi } \delta M-64 \sqrt{\Theta }\right)}+5 \sqrt{\pi } \delta M-32 \sqrt{\Theta }\right)   \right.\right. \\
& \left.\left.   +226688 \sqrt{\pi } \delta M^2 \Theta ^{3/2} \ln \left(3 \sqrt[4]{\pi } \sqrt{\delta M \left(9 \sqrt{\pi } \delta M-64 \sqrt{\Theta }\right)}+9 \sqrt{\pi } \delta M-32 \sqrt{\Theta }\right)   \right.\right. \\
& \left.\left.   -1813504\delta M \Theta ^2 \ln \left(3 \sqrt[4]{\pi } \sqrt{\delta M \left(9 \sqrt{\pi } \delta M-64 \sqrt{\Theta }\right)}+9 \sqrt{\pi } \delta M-32 \sqrt{\Theta }\right)   \right.\right. \\
& \left.\left.    +404352 \sqrt{\pi } \delta M^2 \Theta ^{3/2} \ln \left(\sqrt{\pi } \delta M-8 \sqrt{\Theta }\right)-3234816 \delta M \Theta ^2 \ln \left(\sqrt{\pi } \delta M-8 \sqrt{\Theta }\right)    \right.\right. \\
&\left. \left.     +186624 \sqrt{\pi } \delta M^2 \Theta ^{3/2} \ln \left(\sqrt[4]{\pi } \sqrt{M \left(\sqrt{\pi } \delta M-8 \sqrt{\Theta }\right)}+\sqrt{\pi } \delta M-4 \sqrt{\Theta }\right)    \right.\right. \\
& \left.\left.    -1492992 \delta M \Theta ^2 \ln \left(\sqrt[4]{\pi } \sqrt{\delta M \left(\sqrt{\pi } \delta M-8 \sqrt{\Theta }\right)}+\sqrt{\pi } \delta M-4 \sqrt{\Theta }\right)   \right.\right. \\
& \left.\left.   +226688 \sqrt{\pi } \delta M^2 \Theta ^{3/2} \ln \left(9 \sqrt{\pi } \delta M^2-68 \delta M \sqrt{\Theta }+3 \sqrt{\delta M \left(\sqrt{\pi } \delta M-8 \sqrt{\Theta }\right)} \sqrt{\delta M \left(9 \sqrt{\pi } \delta M-64 \sqrt{\Theta }\right)}\right)   \right.\right. \\
& \left.\left.    -1813504 \delta M \Theta ^2 \ln \left(9 \sqrt{\pi } \delta M^2-68 \delta M \sqrt{\Theta }+3 \sqrt{\delta M \left(\sqrt{\pi } \delta M-8 \sqrt{\Theta }\right)} \sqrt{\delta M \left(9 \sqrt{\pi } \delta M-64 \sqrt{\Theta }\right)}\right)              \right] \right\} .
\end{split}
\fe
It is important to mention that the notation of $\delta M$ means the application of the limits of integration, i.e., $\delta M = M_{rem} - M_{i}$.

%%%%%%%%%%%%%%%%%%%%%%%%%%%%%%%%%%%%%%%%%%%%%%%%%%%%%%%%%%%%%%%%%%%%%%%%%%%%%%%%%%%%%%%%%%%%%%%%%%%%%%%%%%%%%%%%%%%%%%%%%%%%%%%%%%%%%%%%%%%%%%%%%%%%%%%%%%%%%%%%%%%%%%%%%%%%%%%%%%%%%%%%%%%%%%%%%%%%%%%%%%%%%%%%%%%%%%%%%%%%%%%%%%%%%%%%%%%%%%%%%%%%%%%%%%%%%%%%%%%%%%%%%%%%%%%%%%%%%%%%%%%%%%%%%%%%%%%%%%%%%%%%%%%%%%%%%%%%%%%%%%%%%%%%%%%%%%%%%%%%%%%%%%%%%%%%%%%%%%%%%%%%%%%%%%%%%%%%%%%%%%%%%%%%%%%%%%%%%%%%%%%%%%%%%%%%%%%%%%%%

\section*{Acknowledgments}
\hspace{0.5cm}The authors would like to thank the Conselho Nacional de Desenvolvimento Cient\'{\i}fico e Tecnol\'{o}gico (CNPq) for financial support. P. J. Porf\'{\i}rio would like to acknowledge the Brazilian agency CNPQ, grant No. 307628/2022-1. The work by A. Yu. Petrov has been partially supported by the CNPq project No. 303777/2023-0. Moreover, A. A. Araújo Filho is supported by Conselho Nacional de Desenvolvimento Cient\'{\i}fico e Tecnol\'{o}gico (CNPq) and Fundação de Apoio à Pesquisa do Estado da Paraíba (FAPESQ), project No. 150891/2023-7. A. {\"O}. would like to acknowledge the contribution of the COST Action CA21106 - COSMIC WISPers in the Dark Universe: Theory, astrophysics and experiments (CosmicWISPers) and the COST Action CA22113 - Fundamental challenges in theoretical physics (THEORY-CHALLENGES). We also thank TUBITAK and SCOAP3 for their support.

\bibliographystyle{ieeetr}
\bibliography{main}

\end{document}